%% file: arxiv.tex
\pdfoutput=1
\documentclass[journal=acscii,layout=twocolumn, manuscript=article]{achemso}

\usepackage[english]{babel}
\usepackage[utf8]{inputenc}
\usepackage[colorinlistoftodos, color=green!40, prependcaption]{todonotes}
\usepackage{hyperref}
\addto{\captionsenglish}{} 

\input{preamble} 

\author{Simon Axelrod}
    \affiliation{Department of Chemistry and Chemical Biology,
    Harvard University, Cambridge, MA, 02138}
    \alsoaffiliation{Department of Materials Science and Engineering, Massachusetts Institute of Technology, Cambridge, MA, 02139}

\author{Eugene Shakhnovich}
    \affiliation{Department of Chemistry and Chemical Biology, 
    Harvard University, Cambridge, MA, 02138}
  
\author{Rafael Gómez-Bombarelli}
    \email{rafagb@mit.edu}
    \affiliation{Department of Materials Science and Engineering, Massachusetts Institute of Technology, Cambridge, MA, 02139}

\date{\today} 

\title[]{Thermal half-lives of azobenzene derivatives: virtual screening based on intersystem crossing using a machine learning potential }

\begin{document}

\begin{abstract}
Molecular photoswitches are the foundation of light-activated drugs. A key photoswitch is azobenzene, which exhibits \textit{trans}-\textit{cis} isomerism in response to light. The thermal half-life of the \textit{cis} isomer is of crucial importance, since it controls the duration of the light-induced biological effect. Here we introduce a computational tool for predicting the thermal half-lives of azobenzene derivatives. Our automated approach uses a fast and accurate machine learning potential trained on quantum chemistry data. Building on well-established earlier evidence, we argue that thermal isomerization proceeds through rotation mediated by intersystem crossing, and incorporate this mechanism into our automated workflow. We use our approach to predict the thermal half-lives of 19,000 azobenzene derivatives. We explore trends and tradeoffs between barriers and absorption wavelengths, and open-source our data and software to accelerate research in photopharmacology.

\end{abstract}


\maketitle

\section*{Introduction}
Photoswitches are compounds whose properties can be modified by light. They have applications in many developing technologies, such as organic electronics \cite{orgiu201425th}, energy storage \cite{kolpak2011azobenzene}, and targeted medicine \cite{broichhagen2015roadmap}. The latter includes photopharmacology, the field of light-activated drugs. Such drugs are built around photoswitchable scaffolds, which allows their medicinal activity to be controlled with light.  The most common scaffold is azobenzene, which undergoes \textit{cis} $\leftrightarrow$ \textit{trans} isomerization in response to light. An inactive drug built around azobenzene can be activated with light at certain times or in certain regions of the body. This can minimize off-target activity, thereby minimizing side effects \cite{broichhagen2015roadmap}. 
\begin{figure*}[!t]
	\centering
	\includegraphics[width=\textwidth]{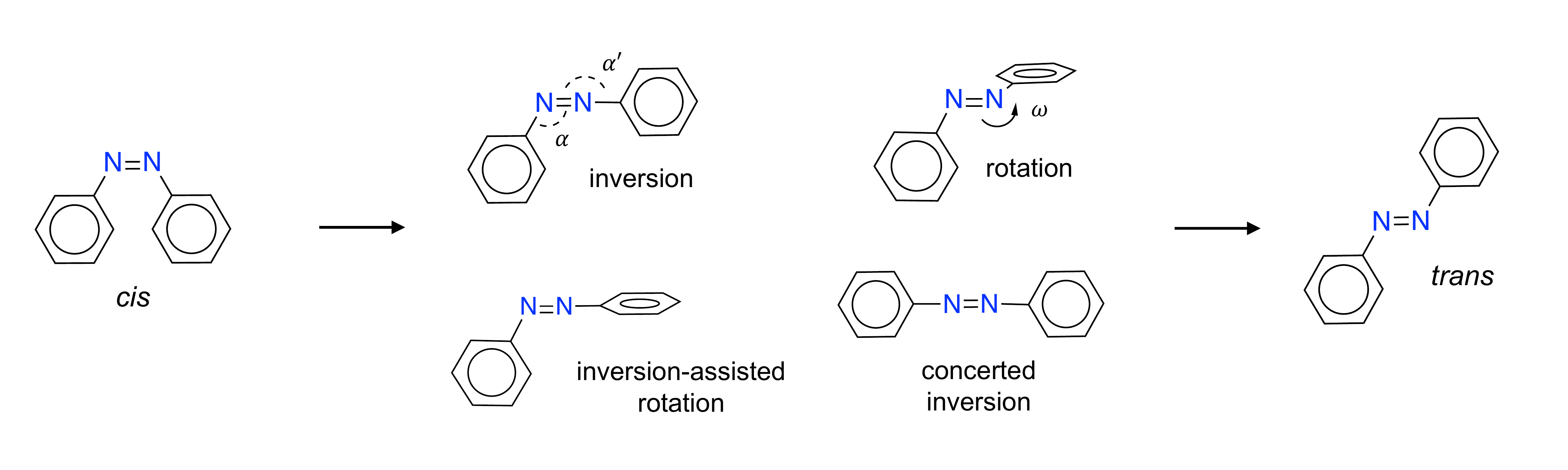}
	\caption{Possible mechanisms of thermal isomerization. The inversion TS has $\alpha$ or $\alpha' \approx 180^{\circ}$, while the rotational TS has $\omega \approx 90^{\circ}$. Inversion-assisted rotation combines inversion and rotation. Concerted inversion combines $\alpha$ inversion and $\alpha'$ inversion.  }
	\label{fig:mechanisms}
\end{figure*}

The development of photoactive drugs is a complex, multi-objective optimization problem. A formidable number of properties must be optimized for all drugs.  Photoactive compounds must also absorb light at the right wavelength, typically in the near infrared region; isomerize with high efficiency in the excited state; display differential bioactivity between the two isomers; and thermally revert to the stable isomer in a specific time frame \cite{lerch2016emerging}. For many applications this time frame should be as long as possible, and so the isomerization barrier should be as high as possible \cite{lerch2016emerging}. Yet substituents that shift absorption from the UV to the visible or near-IR regions often lower the thermal barrier \cite{broichhagen2015roadmap}. This highlights the challenge of the optimization problem.

The design of photoactive drugs can be accelerated with computational modeling.  Property predictors can be applied to large virtual libraries, and the results can be used to narrow the search space of promising compounds \cite{pyzer2015high}. Quantum chemistry can predict many properties with good accuracy, but the calculations are quite slow. In our previous work, we showed how a machine learning (ML) potential trained on quantum chemistry data can be used to rapidly predict the quantum yield of azobenzene derivatives \cite{axelrod2022excited}.  In this work we develop an ML-based computational workflow to predict the thermal half-lives of azobenzene derivatives.

Our contributions are as follows. First, we improve the theory of thermal azobenzene isomerization. In particular, we provide evidence that thermal isomerization proceeds through intersystem crossing, not through a typical singlet transition state (TS). This builds on the theory that was proposed nearly twenty years ago \cite{cembran2004mechanism}, but which has largely been overlooked (with some exceptions \cite{heindl2020rational, kuntze2022towards}).  We also demonstrate the critical importance of multireference effects in barrier calculations.

Second, we provide a fast and user-friendly computational tool for predicting the barriers and absorption wavelengths of azobenzene derivatives. Our program can be run with a single command. The only user input required is the SMILES strings of the relevant compounds. These can be generated programatically, or with visual interface programs such as ChemDraw. Further, the program is quite fast, replicating the results of spin-flip, time-dependent density functional theory (SF-TDDFT)  \cite{shao2003spin} in milliseconds through use of a transferable ML potential. This program adds to the growing collection of computational models for predicting photoswitch properties \cite{axelrod2022excited, adrion2021benchmarking, mukadum2021efficient, griffiths2022data}. Our software and pre-trained models are freely available at \url{https://github.com/learningmatter-mit/azo_barriers}. 

Third, we use our tool to perform virtual screening of nearly 19,000 hypothetical azobenzene derivatives. We identify species with high isomerization barriers and redshifted absorption spectra. The data is freely available at \url{https://doi.org/10.18126/unc8-336t} through the Materials Data Facility \cite{blaiszik2016materials, blaiszik2019data}. Researchers can use the species with favorable properties as scaffolds for new photoactive drugs. Further, we explain these results in terms of substitution patterns and substituent properties. These insights will accelerate the design of meta-stable, red-shifted azobenzene derivatives in the future.

\section*{Theory and methods}
\subsection*{Isomerization mechanisms}
Four mechanisms have been proposed for azobenzene isomerization \cite{bandara2012photoisomerization}: rotation, inversion, inversion-assisted rotation, and concerted inversion (Fig. \ref{fig:mechanisms}). Rotation is characterized by $\omega \approx 90^{\circ}$ and $\alpha \approx \alpha ' \approx 120^{\circ}$. Both inversion mechanisms have $\alpha \approx 180^{\circ}$ and $\alpha ' \approx 120^{\circ}$; pure inversion has $\omega \approx 180^{\circ}$, while inversion-assisted rotation has $\omega \approx 90^{\circ}$. (Some works refer to inversion-assisted rotation simply as rotation \cite{adrion2021benchmarking}, but we avoid that terminology here.) Concerted inversion has $\alpha \approx \alpha ' \approx 180^{\circ}$, but can be excluded from possible thermal mechanisms; see  Supporting Information (SI) Sec. \ref{subsec:concerted}. Here we group together inversion and inversion-assisted rotation, and refer to both as inversion (SI Sec. \ref{subsec:inversion}). For asymmetrically substituted azobenzenes, $\alpha=180^{\circ}$ is distinct from $\alpha'=180^{\circ}$, and $\omega=90^{\circ}$ is distinct from $\omega=-90^{\circ}$. This gives two inversion TSs and two rotational TSs.

\subsection*{Standard models}

Several works have predicted the thermal lifetimes of azobenzene derivatives with computational methods \cite{dokic2009quantum, knie2014ortho, rietze2017thermal, schweighauser2015attraction, adrion2021benchmarking, mukadum2021efficient}. However, there are two main issues with previous calculations. First, all levels of theory overestimate the experimental enthalpy and entropy of activation. The experimental activation entropy is $-50.2$ J $\mathrm{mol}^{-1} \mathrm{K}^{-1}$ \cite{asano1981temperature}, while Hartree-Fock, MP2, CC2, and DFT with 12 different functionals predict values between $+7$ and $+28$ J $\mathrm{mol}^{-1} \mathrm{K}^{-1}$ \cite{rietze2017thermal}. At room temperature, the experimental entropic contribution to $\Delta G^{\dagger}$ is then $-T \Delta S^{\dagger} = +3.6$ kcal/mol, while the computational contribution is between $-0.5$ and $-2$ kcal/mol. Most of the error persists even after corrections to the harmonic approximation \cite{rietze2017thermal}.

Each method also overestimates $\Delta H^{\dagger}$. The experimental activation enthalpy is $21.1$ kcal/mol \cite{asano1981temperature}; B3LYP-D3/6-311++G** predicts $25.2$ kcal/mol, and CC2/aug-cc-pVTZ predicts 29.8 kcal/mol \cite{rietze2017thermal}. CASPT2(10,8)/6-31G* gives $\Delta E^{\dagger} = 31.0$ kcal/mol \cite{casellas2016excited}, which is nearly identical to CC2/aug-cc-pVTZ \cite{rietze2017thermal}. These calculations, like most in the literature, were performed for the inversion TS. In SI Sec. \ref{sec:benchmark}, we show that highly accurate spin-flip coupled cluster methods yield similar overestimates for the rotational TS.

The errors in $\Delta H^{\dagger}$ and $-T \Delta S^{\dagger}$ partially cancel for $\Delta G^{\dagger}$. 
Indeed, recent work has extensively benchmarked different levels of theory for predicting $\Delta G^{\dagger}$ of various azoarenes, and found errors near 1 kcal/mol for B3LYP-D3 with some basis sets \cite{adrion2021benchmarking}. However, given the significant error cancellation for $\Delta G^{\dagger}$, and the underestimation of $\Delta H^{\dagger}$ relative to correlated wavefunction methods, these results should be interpreted with caution.

The second issue concerns the rotational mechanism. While most DFT calculations have been applied to inversion \cite{casellas2016excited, rietze2017thermal, adrion2021benchmarking}, CASPT2 calculations indicate that rotation is in fact preferred for azobenzene \cite{cembran2004mechanism}. 
Yet Ref. \cite{rietze2017thermal} found that the rotational TS cannot be optimized with B3LYP, a finding that we have also reproduced for various derivatives. 
It is troubling that rotation can be energetically favored, yet the rotational TS cannot be reached through DFT optimization. 

\subsection*{Intersystem crossing}
\begin{figure}[!t]
	\centering
	\includegraphics[width=0.95\columnwidth]{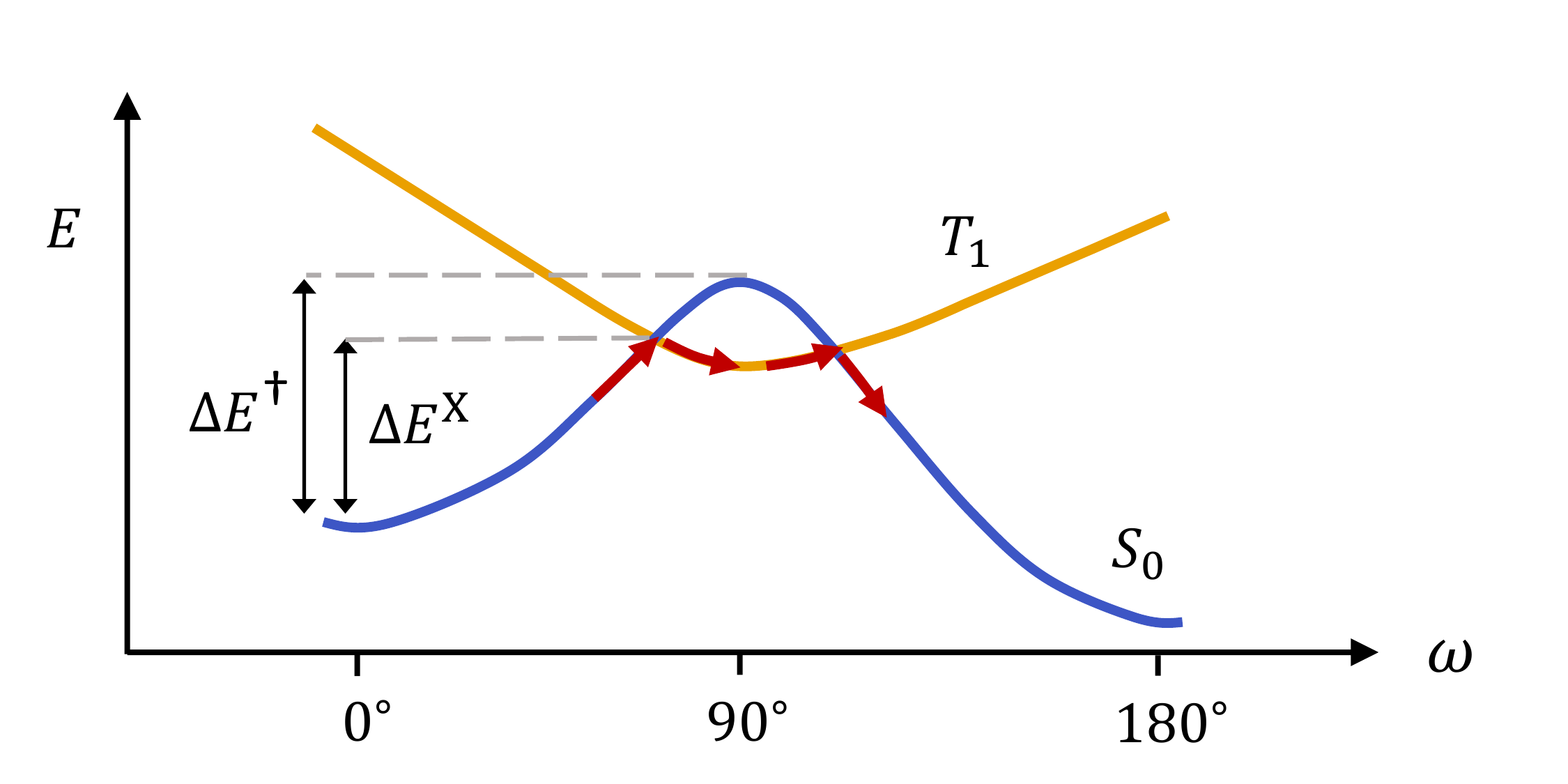}
	\caption{Schematic of triplet-mediated thermal isomerization. Intersystem crossing from singlet ($S_0$) to triplet ($T_1$) occurs near $\omega = 70^{\circ}$. The energy is lower at the crossing point than at the $S_0$ TS ($\Delta E^{\mathrm{X}} < \Delta E^{\dagger}$). Re-crossing from $T_1$ to $S_0$ then occurs near $\omega = 105^{\circ}$, and isomerization is completed on the $S_0$ surface.}
	\label{fig:isc}
\end{figure}
\begin{figure*}[!t]
	\centering
	\includegraphics[width=\linewidth]{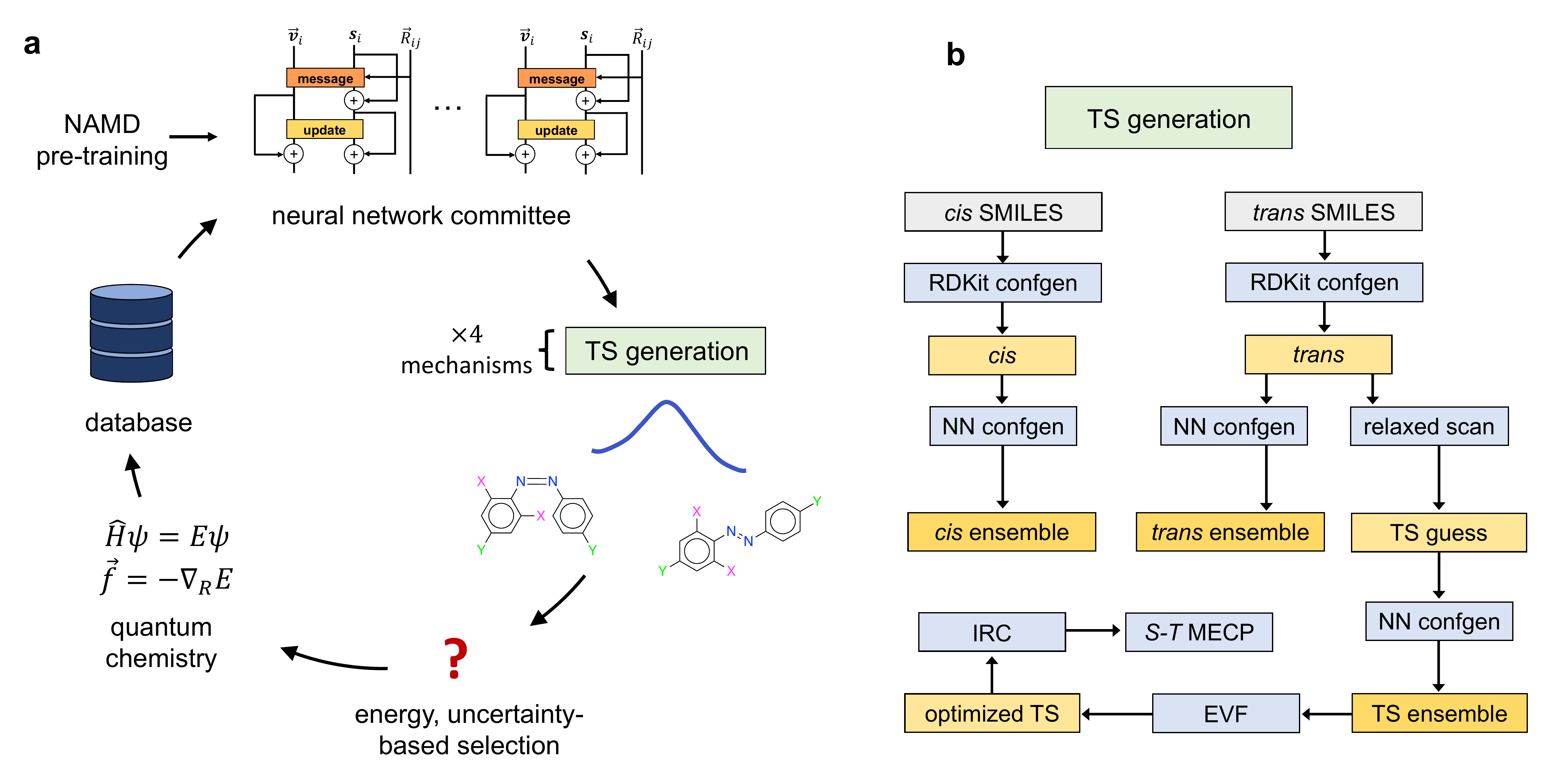}
	\caption{Approach to active learning and TS generation used in this work. (a) Active learning loop for training the NN. (b) Workflow for generating equilibrium and TS geometries. ``confgen'' stands for conformer generation, which is described in SI Sec. \ref{sec:confgen}. ``$S$-$T$ MECP'' denotes the search for the minimum-energy singlet-triplet crossings on each side of the TS.}
	\label{fig:ts_and_al}
\end{figure*}
To address the overestimation of $\Delta S^{\dagger}$ and $\Delta H^{\dagger}$ common to all levels of theory, we advocate intersystem crossing (ISC) as the mechanism of thermal isomerization (Fig. \ref{fig:isc}). This was proposed in Ref. \cite{cembran2004mechanism} nearly 20 years ago. As discussed below, $\Delta G^{\dagger}$ is replaced with $\Delta G^{\mathrm{X}}$, the free energy difference at the singlet-triplet crossing geometry $\mathrm{X}$. The rate prefactor $k_{\mathrm{B}} T / h$ is replaced with the ISC rate $k_{\mathrm{ISC}}$. Since the crossing geometry has a lower energy than the TS, this approach corrects the overestimation of $\Delta H^{\dagger}$. Further, the measured $\Delta S^{\dagger}$ is not a true activation entropy, but in fact related to both $\mathrm{log}[k_{\mathrm{ISC}} / (k_{\mathrm{B}} T / h)]$ and  $\Delta S^{\mathrm{X}}$ (SI Eq. (\ref{eq:s_app})). The negative value of $\Delta S^{\dagger}$ partly reflects the fact that $k_{\mathrm{ISC}} < k_{\mathrm{B}} T / h$. Computing $k_{\mathrm{ISC}}$ with CASPT2(14,12)/6-31G* yields good agreement with the effective experimental activation entropy, $\Delta S^{\mathrm{eff}}$, for unsubstituted azobenzene \cite{cembran2004mechanism}.

We note, however, that isomerization can still proceed through $S_0$ for certain derivatives and environments, and that the $S_0$ rate should be compared to the triplet-mediated rate for any new system. Indeed, Ref. \cite{liang2022multiscale} found good agreement between the $S_0$ CASPT2 rate and the experimental rate for phototrexate in DHFR, though no comparison was made to the experimental activation entropy. For this reason we compute the rate with both Eyring TS theory and ISC for all molecules in this work.  
%
%

\subsection*{Multireference effects}
To properly optimize the rotational TS, we show that it is critical to include multi-reference effects. This is not surprising, since there is a conical intersection when $\omega$ is close to $90^{\circ}$ \cite{yue2018performance}. In this work we use SF-TDDFT \cite{shao2003spin}, since it accounts for some double excitations and generally provides an accurate description of conical intersections \cite{lee2019conical}. We use the common BHHLYP functional \cite{becke1993new} and 6-31G* basis \cite{francl1982self}. As shown in SI Sec. \ref{sec:benchmark}, single-reference methods produce rotational TS cusps. This is the likely cause of the failed rotational TS optimizations with standard DFT. Multi-reference methods, by contrast, produce smooth maxima. We note that the inversion TS is \textit{also} close to a conical intersection \cite{cembran2004mechanism}, which further reinforces the need for a multi-reference treatment of azobenzene TSs.

SF-TDDFT accounts for multi-reference effects while offering a reasonable balance between cost and accuracy. Empirically it has sub-cubic scaling \cite{axelrod2022excited}, which makes it far more affordable than highly accurate spin-flip coupled cluster methods that scale as $N^7$ \cite{manohar2008noniterative}. Further, as shown in SI Sec. \ref{sec:benchmark}, SF-TDDFT has similar errors to CASPT2 for rotational barriers in azobenzene. The latter scales as $N^5$ for a fixed active space, requires manual active space selection, and does not have analytic gradients in most quantum chemistry packages, which are essential for training ML potentials. While SF-TDDFT is not spin-complete, we have found that $S_0$ spin contamination is rather low. As discussed in SI Sec. \ref{subsec:training_singlet}, the average square spin in the $S_0$ state is only 0.16, and we excluded all data with square spin exceeding 1.0. Lastly, the errors in SF-TDDFT are largely systematic, as demonstrated by the strong correlation between predicted and experimental activation free energies in Fig. \ref{fig:barriers_vs_expt}. This is discussed further in the Results section below.

\subsection*{Computational workflow}
Standard quantum chemistry approaches are rather slow. To address this and the above issues, we develop an ISC workflow based on ML potentials \cite{axelrod2022learning} that are trained on multi-reference SF-TDDFT data.

We generate initial TSs through a relaxed scan, and refine the structures with a conformer search \cite{pracht2020automated, 2020geom, axelrod2020molecular} and eigenvector following (EVF, Fig. \ref{fig:ts_and_al}(b)).  We then use the intrinsic reaction coordinate (IRC) \cite{ishida1977intrinsic} to locate singlet-triplet crossings on either side of each rotational TS. The geometries are subsequently refined with a minimum energy crossing point (MECP) optimization (SI Sec. \ref{sm_sec:mecp}). We also apply Eyring TS theory to the TSs and compare the results to the ISC approach.

Four relaxed scans are performed for each of the four mechanisms (two inversion and two rotation). Conformers are generated for each TS guess using a conformational search with fixed CNNC atoms. The five lowest-energy conformers for each mechanism are then optimized with eigenvector following, yielding 20 TSs. The TS with the lowest free energy for each rotational mechanism is used to find the singlet-triplet crossings. ISC does not occur for the inversion mechanism \cite{cembran2004mechanism}.

Once the workflow is completed, the ISC-based reaction rate is calculated as 
\begin{align}
	k_{\mathrm{X}} = k_{\mathrm{ISC}} \mathspace \mathrm{exp}\left( - \Delta G^{\mathrm{X}} / k_{\mathrm{B}} T  \right),
\end{align}
where $G$ is given in SI Eqs. (\ref{eq:g_total})-(\ref{eq:s_conf}). $\Delta G^{\mathrm{X}}$ depends on the energies, conformer ensembles, and vibrational frequencies of both the reactant and the crossing geometry. We compute $k_{\mathrm{ISC}}$ using the approach in Ref. \cite{liu2014modeling}, as described in SI Sec. \ref{sm_sec:na_tst}. The expression depends on the spin-orbit coupling, temperature, and forces on the singlet and triplet surfaces at the crossing. The spin-orbit coupling is taken as constant, $H_{\mathrm{SO}} \approx 20 \ \mathrm{cm}^{-1}$, as described in SI Sec. \ref{sm_sec:na_tst}. The coupling is about 20 times larger than the typical value for planar aromatic compounds, because of the $n-\pi^*$ character of the triplet state \cite{cembran2004mechanism}. This corresponds to a 400-fold enhancement in the ISC rate. 

We also compute the reaction rate from Eyring TS theory, given by
\begin{align}
	k_{\mathrm{TST}} = \frac{ k_{\mathrm{B}} T }{h} \mathrm{exp} \left(- \Delta G^{\dagger} / k_{\mathrm{B}} T \right). \label{eq:tst}
\end{align}
For ease of comparison we convert ISC-based reaction rates into the form of Eq. (\ref{eq:tst}), where $\Delta G^{\dagger}$ is replaced by the effective activation free energy,
\begin{align}
    & \Delta G^{\mathrm{eff}} = -k_{\mathrm{B}} T \ \mathrm{log}\left( \frac{h k_{\mathrm{X}}}{k_{\mathrm{B}} T} \right) \nonumber \\
    & = \Delta H^{\mathrm{X}} - T\Delta S^{\mathrm{eff}} - \frac{1}{2} k_{\mathrm{B}} T.
\end{align}
$\Delta S^{\mathrm{eff}}$ depends on both $k_{\mathrm{ISC}}$ and $\Delta S^{\mathrm{X}}$, and is given in SI Eq. (\ref{eq:s_app}).

\begin{table*}[t]
\small
\centering
\begin{tabular}{c|c|c|c|c|c}
    \hline
    Geometry type & Model type & \ \begin{tabular}{@{}c@{}}Singlet $\Delta{E}$\\ error\hyperlink{a}{\textsuperscript{a}} \end{tabular} \ & \ \begin{tabular}{@{}c@{}}Triplet $\Delta{E}$ \\ error\hyperlink{b}{\textsuperscript{b}} \end{tabular} \ & \ \begin{tabular}{@{}c@{}}Singlet $\vec{F}$ \\ error \end{tabular} \ & \  \begin{tabular}{@{}c@{}}Triplet $\Delta \vec{F}$ \\ error\hyperlink{c}{\textsuperscript{c}}  \end{tabular}  \ \\
    \hline
    & One model & 0.81 & 0.23 & 0.44 & 0.48 \\ 
    \ Optimized TS\hyperlink{d}{\textsuperscript{d}} \ & Ensemble\hyperlink{e}{\textsuperscript{e}} & 0.73 & 0.19 & 0.34 & 0.44 \\ 
    & \ Ensemble, lowest $95\%$ uncertainty\hyperlink{f}{\textsuperscript{f}} 
    \ & 0.66 & 0.16 & 0.31 & 0.42 \\
    \hline
    & One model & 1.09 &  0.34 & 0.69 & 0.55 \\ 
    \ TS metadynamics\hyperlink{g}{\textsuperscript{g}} \ &  Ensemble  & 0.99 & 0.31 & 0.54 & 0.50 \\
    & \ \ Ensemble, lowest $95\%$ uncertainty \ & 0.86 & 0.27 & 0.49 & 0.48 \\
    \hline
\end{tabular}
\caption{Model performance for 334 species outside the training set. Units are kcal/mol for energies and kcal/mol/\AA \ for forces. Forces are denoted by $\vec{F}$. }
\label{tab:model_accuracy}
\flushleft
\footnotesize{\hypertarget{a}{\textsuperscript{a}} Singlet $\Delta E = E - E_{\mathrm{cis}}$, where $E_{\mathrm{cis}}$ is the energy of the lowest energy \textit{cis} conformer. \\
\hypertarget{b}{\textsuperscript{b}} Triplet $\Delta E = E_{\mathrm{S}} - E_{\mathrm{T}}$, where $E_{\mathrm{S}} $ is the singlet energy and $E_{\mathrm{T}} $ is the triplet energy. \\
\hypertarget{c}{\textsuperscript{c}} Triplet $\Delta \vec{F} = \vec{F}_{\mathrm{S}} - \vec{F}_{\mathrm{T}}$. \\
\hypertarget{d}{\textsuperscript{d}} Four TSs per species, one for each mechanism. 
} \\
\hypertarget{e}{\textsuperscript{e}} Three models. \\
\hypertarget{f}{\textsuperscript{f}} Uncertainty computed as the standard deviation of the three model predictions. \\
\hypertarget{g}{\textsuperscript{g}} Geometries randomly sampled from NN metadynamics for TS conformer generation. Five geometries were sampled for each species.
\end{table*}
\subsection*{ML models}
We train separate models to predict the $S_0$ energy, $T_1$ energy, and $S_0/S_1$ gap. Our models use the PaiNN architecture \cite{schutt2021equivariant}, which predicts molecular properties through equivariant message-passing. This approach generates a feature vector for each atom that incorporates information from its surrounding environment. The initial feature vector is generated from the atomic number alone, and is then updated through a set of ``messages''. These messages incorporate the distance, orientation, and features of atoms within a cutoff distance. The messages are then used to update the atomic feature vectors. This is performed several times, which leads to information being combined from increasingly distant atoms. Lastly, the atomic features are mapped to per-atom energies using a neural network, which are summed to yield the molecular energy. The forces are computed through automatic differentiation of the energy.

We pre-train the models on 680,736 gas-phase SF/6-31G* calculations from non-adiabiatc molecular dynamics (NAMD), which were previously generated in Ref. \cite{axelrod2022excited}. We then refine the models using approximately 40,000 SF/6-31G* calculations with a C-PCM model of water \cite{truong1995new, barone1998quantum, cossi2003energies}. Pre-training on existing gas-phase data means that fewer new solvent calculations are required to reach a target accuracy \cite{ang2021active}.

The geometries for SF-TDDFT/C-PCM calculations are generated through active learning \cite{ang2021active, wang2020active, axelrod2022excited} based on our TS workflow (Fig. \ref{fig:ts_and_al}). In each round of active learning, we train three $S_0$ models on previous SF-TDDFT/C-PCM data (the gas-phase model is used in the first round). The difference in model predictions is used to identify geometries that are poorly described by the model. These high-uncertainty configurations, together with some geometries that are sampled randomly or by energy, receive new quantum chemistry calculations (see SI Sec. \ref{sec:si_ex_methods}).

\section*{Results}
\subsection*{Model performance}
The model accuracy is shown in Table \ref{tab:model_accuracy}. Mean absolute errors (MAEs) are given for the singlet and triplet models, for both optimized TSs and off-equilibrium geometries sampled during TS conformer generation (see SI Sec. \ref{sec:confgen}). All geometries come from species outside the training set.

The model performance is excellent. The error in the barrier energy, $\Delta E = E_{\mathrm{TS}} - E_{\mathrm{cis}}$, is 0.81 kcal/mol for one model and 0.73 kcal/mol for an ensemble of three models. The ensemble error falls to  0.66 kcal/mol after excluding the top 5\% most uncertain geometries. These errors are far below 1.0 kcal/mol, which is the typical definition of chemical accuracy. The model error is significantly smaller than the SF-TDDFT error (SI Sec.  \ref{sec:benchmark}).

The force predictions are similarly accurate, with MAEs below 0.45 and 0.7 kcal/mol/\AA \ for optimized and distorted TSs, respectively.  The performance of the triplet model is even better, with errors that are four times lower than that of the singlet model. Lastly, the $S_0/S_1$ gap model has an MAE of 0.68 and 2.35 kcal/mol for optimized \textit{cis} and \textit{trans} geometries, respectively. These are errors of 3.8 and 13.1 nm for a typical absorption wavelength of 400 nm. 

\subsection*{Comparison to experiment}

\begin{figure*}[t]
	\centering
	\includegraphics[width=\textwidth]{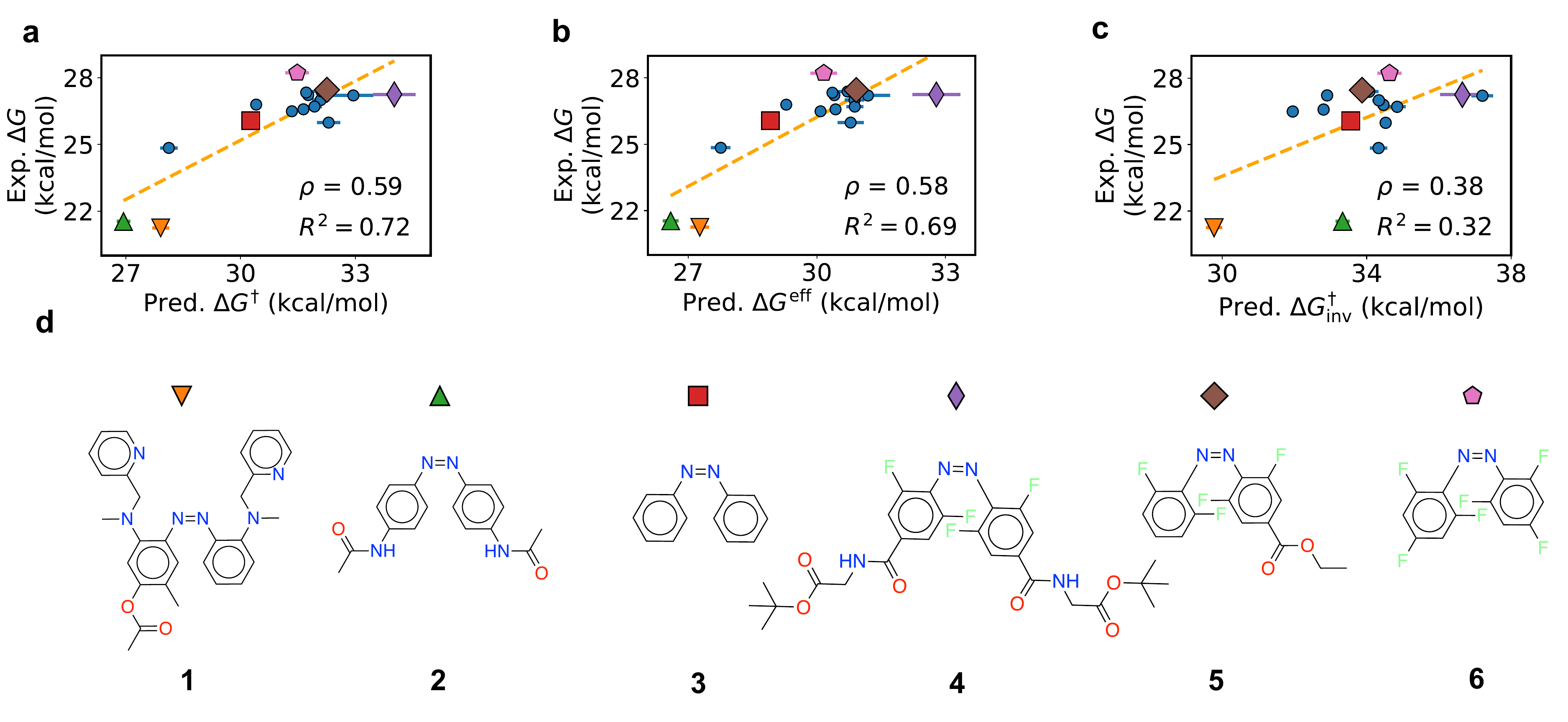}
	\caption{Experimental vs. predicted activation free energies. Dotted orange lines are linear regression results from predicted to experimental values. $\rho$ denotes the Spearman rank correlation. $R^2$ is computed between  the regression results and the experimental data. Error bars are the standard deviation of the energy predictions from three models. (a) Prediction accuracy using TS theory. (b) Prediction accuracy using intersystem crossing. (c) As in (a), but with only the inversion mechanism. (d) Selected compounds highlighted in panels (a)-(c).}
	\label{fig:barriers_vs_expt}
\end{figure*}

The predicted and experimental activation free energies are compared in Fig. \ref{fig:barriers_vs_expt}. The experimental data come from Refs. \cite{gegiou1968temperature, nagamani2005photoinduced, sierocki2006photoisomerization, dokic2009quantum, bandara2010proof,  bandara2011short, beharry2011azobenzene, knie2014ortho, schweighauser2015attraction, moreno2016sensitized}, and can be found in the file containing the virtual screening results. 26 measurements were accessed in total, and 17 of these were used in Fig. \ref{fig:barriers_vs_expt}. We only used the measurements performed in solvent with dielectric constant $\varepsilon \geq 30$. Our model was trained on implicit solvent calculations using $\varepsilon=78.4$ for water. However, we found that the quantum chemistry energies were quite similar when using $\varepsilon=30$, and so included these measurements in the benchmark as well. 

The results of TS and ISC theory are shown in panels (a) and (b), respectively. The rotation mechanism was favored in TS theory for all compounds shown, and so panel (a) equivalently shows $\Delta G^{\dagger}$ from rotation. Panel (c) shows the results of TS theory when considering only the two inversion mechanisms. 

The correlation with experiment is quite good for both TS and ISC theory.  Both have  Spearman rank coefficients $\rho$ near 0.6, and both have $R^2$ near 0.7 after linear regression. The MAEs after regression are 0.79 and  0.85 kcal/mol for TS and ISC theory, respectively. When including all 26 measurements performed in any solvent, we find that $\rho$ actually increases to 0.67 for both methods. $R^2$ falls to 0.58 and 0.63 for TS and ISC theory, respectively. The respective MAEs climb to 0.92 and 0.97 kcal/mol.

The species are properly separated into low-, medium-, and high-barrier groups in Fig. \ref{fig:barriers_vs_expt}. For example, species $\textbf{1}$ and $\textbf{2}$ are predicted to have low barriers, the fluorinated derivatives \textbf{4}-\textbf{6} to have high barriers, and azobenzene to lie in the middle. The models even have some success comparing fluorinated derivatives to each other, with $\rho=0.33$ and $0.42$ among these species for $\Delta G^{\dagger}$ and $\Delta G^{\mathrm{eff}}$, respectively. However, the two approaches give $R^2=0.02$ and $0.04$, respectively. This means that the numerical error is close to that of a random predictor, even though the rankings are better than random.

Both approaches overestimate the barriers, on average by 4.94 and 3.76 kcal/mol for TS theory and ISC, respectively. However, the overestimation is largely systematic, as demonstrated by the high $R^2$ value after linear regression. Further, it is a consequence of SF-TDDFT, not the models. Indeed, as shown in SI Sec. \ref{sec:benchmark}, $\Delta H$ is well-reproduced by the accurate and expensive method SF-EOM-CCSD(dT) \cite{manohar2008noniterative}. SF-TDDFT overestimates $\Delta H$ with respect to both experiment and SF-EOM-CCSD(dT). Transfer learning to this higher level of theory could therefore be of interest in the future.

$\Delta G^{\mathrm{eff}}$ is lower than $\Delta G^{\dagger}$ on average, but otherwise their trends are quite similar. One reason is that rotation is the predicted mechanism for all species. Since each singlet-triplet crossing is on either side of a rotational TS, its energy is correlated with that of the TS. Indeed, the correlation between the two methods is near unity, with $\rho=0.97$ and $R^2=0.98$. This reflects the fact that $E^{\dagger}_{\mathrm{rot}}- E^{\mathrm{X}}$ and $k_{\mathrm{ISC}}$ are nearly constant among different species. However, we explain below that noticeable differences arise when screening large virtual libraries.

The approaches' strong performance should be contrasted with TS theory using only inversion. These results are shown in panel (c). The performance is far worse than using rotation or ISC, with $R^2$ reduced by over 50\%, and $\rho$ reduced by over $35\%$. The MAEs after regression are 1.24 kcal/mol for polar solvents and 1.43 kcal/mol for all data. Note that most works with 
DFT have only produced inversion TSs. This is likely because of failed rotation optimizations, which we attribute to the single-reference nature of DFT and the associated TS cusps (SI Sec. \ref{sec:benchmark}). Similar difficulties were found in Ref. \cite{kuntze2022towards}. Our results highlight the importance of rotation and multi-reference effects.

Note that all species in Fig. \ref{fig:barriers_vs_expt} were in the training set. Hence the comparison to experiment does not measure the model's ability to generalize to new compounds. Rather, it mainly measures the reliability of the workflow and the underlying quantum chemistry. The models' ability to generalize to new species was shown in Table \ref{tab:model_accuracy}.

\subsection*{Virtual screening}
\begin{figure*}[t]	
	\includegraphics[width=\textwidth]{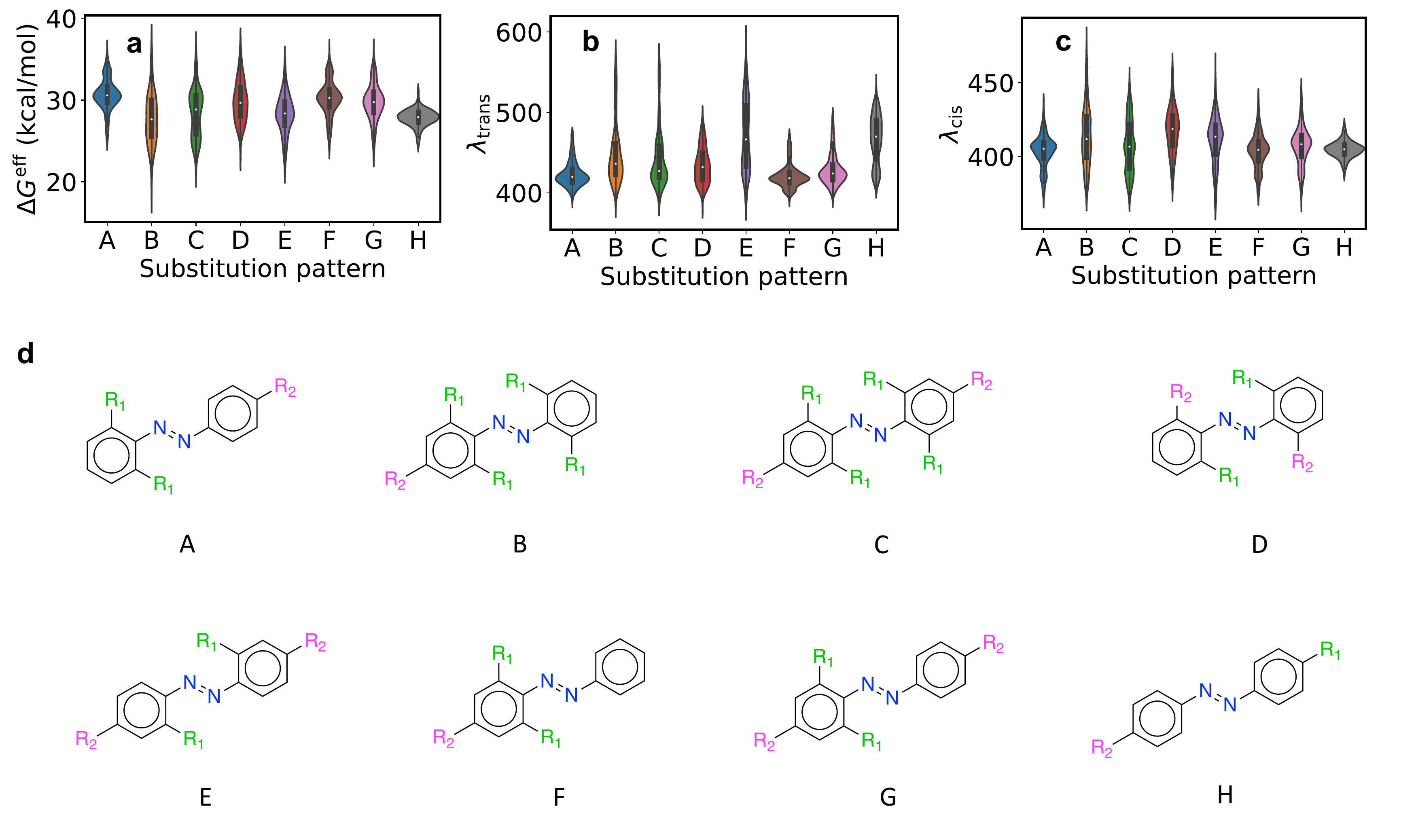}
	\caption{Chemical properties by substitution pattern. 3-$\sigma$ outliers were removed for ease of visualization. (a) Effective activation free energy. (b) \textit{Trans} absorption wavelength. (c) \textit{Cis} absorption wavelength. (d) Substitution patterns.}
	\label{fig:results_by_pattern}
\end{figure*}

With reliable models and predictive workflows, we next screened a virtual library of azobenzene derivatives for key properties in photopharmacology. We ran the workflow of Fig. \ref{fig:ts_and_al}(b) for 25,000 compounds in all. The compounds were generated using the common literature substitution patterns in Fig. \ref{fig:results_by_pattern}(d), folllowing the approach of Refs. \cite{gomez2016design, axelrod2022excited}. The substituents are a combination of literature groups and basic chemical moieties, and can be found with the screening results online. After applying various filters, such as the right number of imaginary frequencies, converged TSs for all mechanisms, and the proper IRC endpoints,  we were left with 19,000 species in total (see SI Sec. \ref{sm_sec:filtering}). 

\subsubsection*{Distributions}
\begin{figure*}[t]
	\centering
	\includegraphics[width=\textwidth]{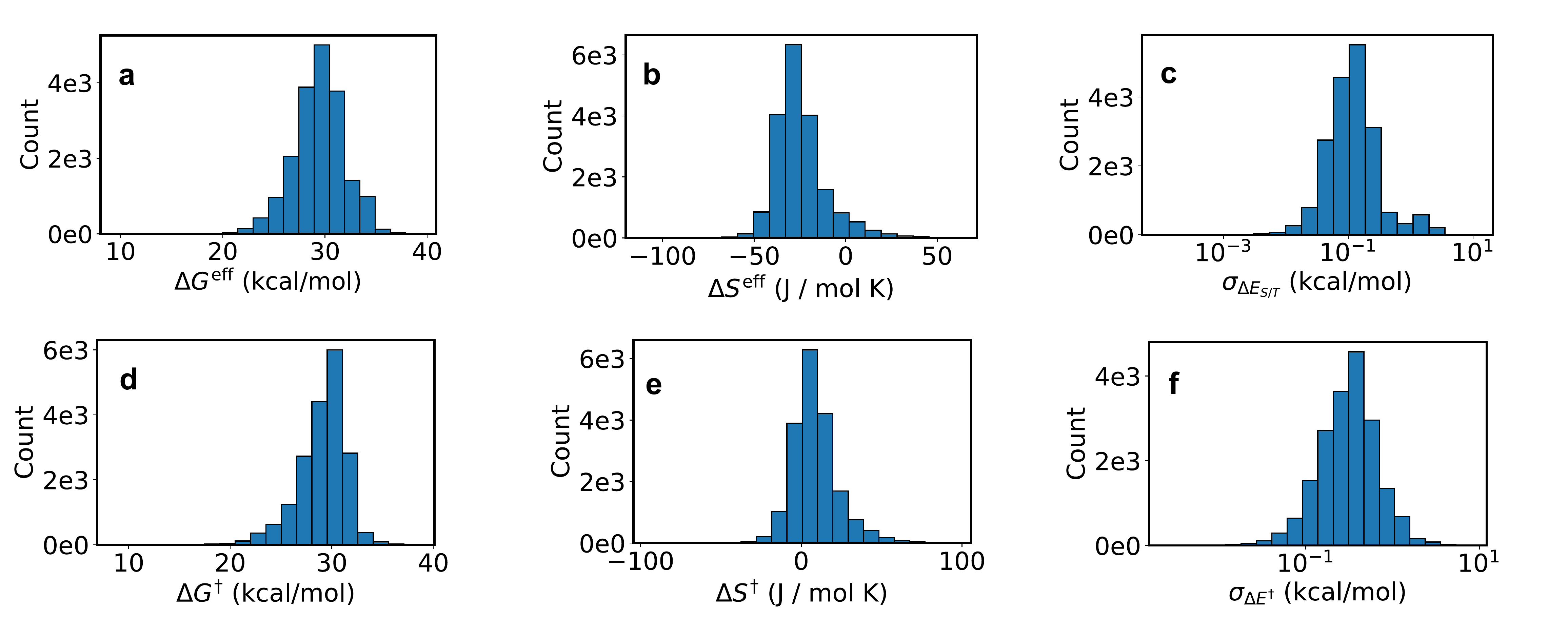}
	\caption{Distribution of various quantities among the 19,000 screened derivatives. (a)  Effective activation free energy. (b) Effective activation entropy. (c) Model uncertainty in the singlet-triplet gap at MECPs. (d)-(e) As in (a)-(b), but using TS theory. (f)  Model uncertainty in the activation energies. }
	\label{fig:general_trends}
\end{figure*}
\begin{figure*}[t]
	\centering
	\includegraphics[width=\textwidth]{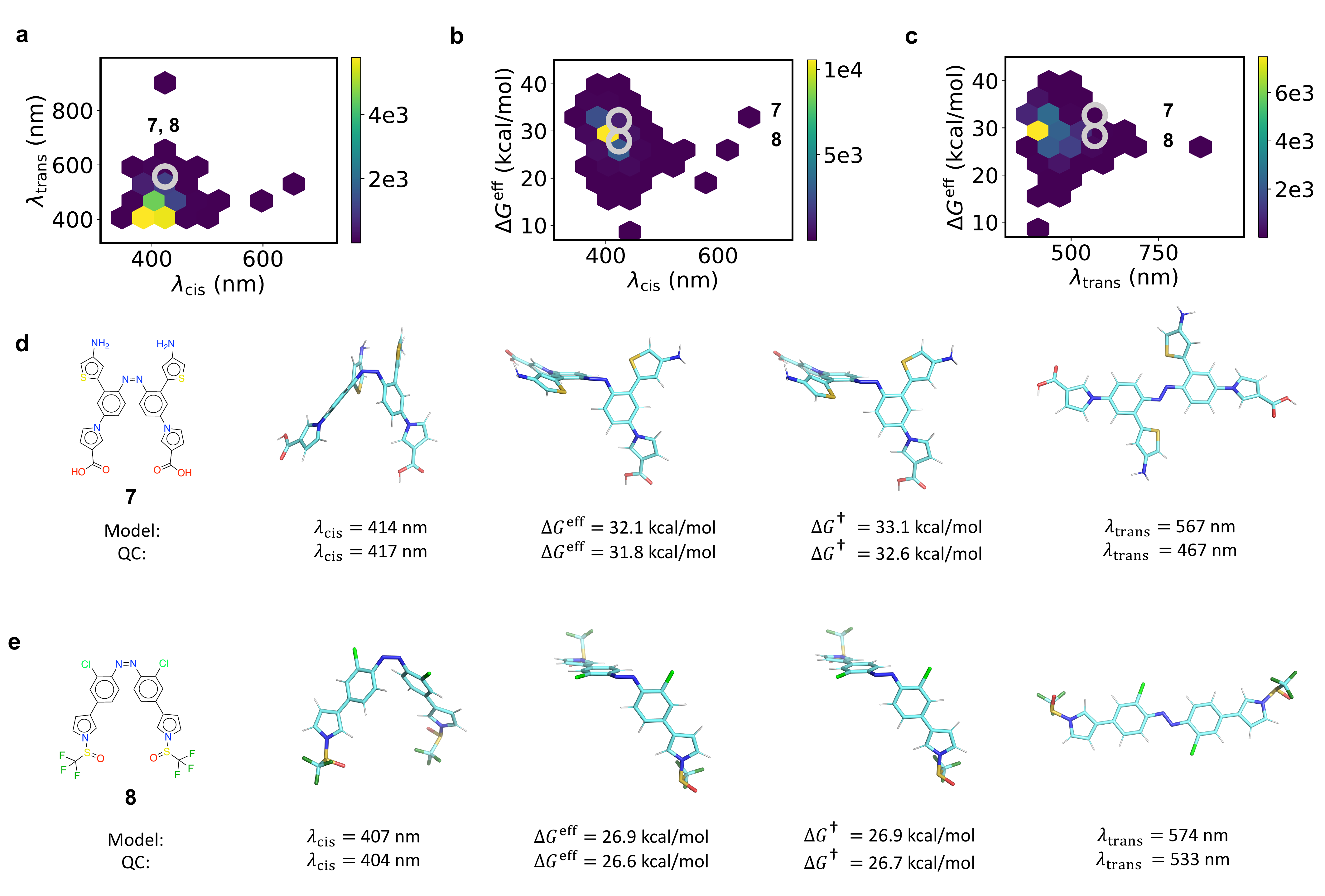}
	\caption{Absorption wavelengths and thermal barriers. (a) \textit{Trans} vs. \textit{cis} and  absorption wavelengths. (b), (c) $\Delta G^{\mathrm{eff}}$ vs. \textit{cis} and \textit{trans} absorption wavelengths, respectively. (d), (e) The two compounds of interest. For each panel the graph is shown on the left, followed by the \textit{cis} geometry, the singlet-triplet crossing closer to \textit{cis}, the TS, and the \textit{trans} geometry. Model and quantum chemistry (QC) predictions are shown below. The compounds are circled in panels (a)-(c). }
	\label{fig:lambda_and_barrier}
\end{figure*}

Figure \ref{fig:general_trends}(a) shows the distribution of $\Delta G^{\mathrm{eff}}$ from this screen. The mean and median are 29.6 and 29.7 kcal/mol, respectively, while unsubstituted azobenzene has a value of 28.9 kcal/mol. The average derivative is thus more kinetically stable than azobenzene. The standard deviation of is 2.6 kcal/mol, which is a factor of 80 in the isomerization rate. 39\% of species (7,400) have a lifetime that is over 10$\times$ that of azobenzene; 19\% (3,600) have a lifetime that is less than $1/10^{\mathrm{th}}$. We conclude that the lifetime is highly tunable using the substitutions in this work.

\begin{figure*}[t]
	\includegraphics[width=\textwidth]{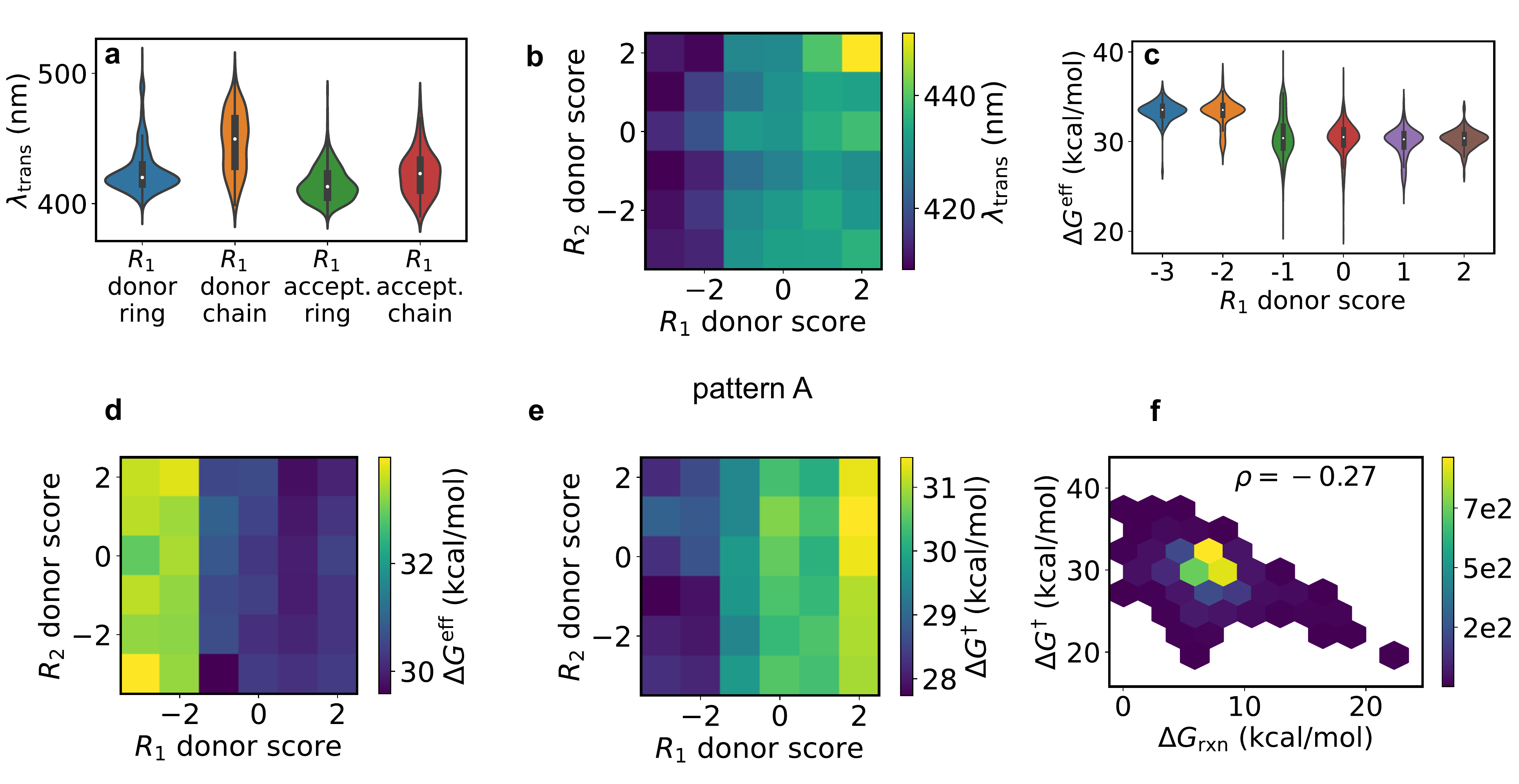}
	\caption{Relationships among substituent properties, free energies, and absorption wavelengths for pattern A. The pattern is shown in Fig. \ref{fig:results_by_pattern} (d). (a) $\lambda_{\mathrm{trans}}$ for donor/acceptor and ring/non-ring $R_1$ substituents. (b) $\lambda_{\mathrm{trans}}$ as a function of $R_1$ and $R_2$ donor scores. (c) $\Delta G^{\mathrm{eff}}$ as a function of $R_1$ donor score. (d) $\Delta G^{\mathrm{eff}}$  as a function of both $R_1$ and $R_2$ donor scores. (e) As in (d), but for $\Delta G^{\dagger}$. (f) $\Delta G^{\mathrm{eff}}$ vs. $\Delta G_{\mathrm{rxn}}$.}
	\label{fig:barrier_trends_A}
\end{figure*}
Panel (b) shows the effective activation entropy. The mean is $-$24.5 J / mol K and the median is $-$26.9 J / mol K. The calculated and experimental values for azobenzene are $-$29.0 J  / mol K and $-$50.2 J / mol K \cite{asano1981temperature}, respectively. The associated error in the entropic free energy is 1.5 kcal/mol. Using the TS approach gives $\Delta S^{\dagger} = 4.7 $ J  / mol K, which has a much higher entropic  free energy error of 3.9 kcal/mol. Few other derivatives have experimental activation entropies, and of those that do, most have values near that of azobenzene. However, of the few with values far from azobenzene, none were predicted accurately by the model \cite{wazzan2010cis, schweighauser2015attraction, moreno2016sensitized}. These cases should be investigated in more detail in the future.

Panels (d) and (e) are analogous to (a) and (b), but with TS theory instead of the ISC approach. Each distribution resembles its partner from ISC theory. However, $\Delta G^{\dagger}$ is more asymmetric than $\Delta G^{\mathrm{eff}}$, with a much steeper drop-off for higher barriers. The reason is as follows. Within TS theory using SF-TDDFT in water, rotation is more often the preferred mechanism (SI Sec. \ref{sm_sec:dg_by_mechanism}). However, inversion can become preferred for species with high enough rotation barriers. This mechanism is not available in ISC theory, since $T_1$ is always higher than $S_0$ during inversion. Hence inversion lowers the high barriers in TS theory, but cannot do the same in ISC theory.

In principle one should calculate both $\Delta G^{\dagger}$ and $\Delta G^{\mathrm{eff}}$, and use the lower value for the reaction rate. If $\Delta G^{\dagger}_{\mathrm{inv}}$ were low enough, it would replace $\Delta G^{\mathrm{eff}}$ in the high barrier regime, and so the steep drop-off of panel (d) would be observed. However, as discussed in SI Sec. \ref{sm_sec:dg_by_mechanism}, SF-TDDFT is not accurate enough to compare absolute $\Delta G^{\dagger}$ and $\Delta G^{\mathrm{eff}}$ directly. Hence a more accurate treatment of this problem would be of interest in the future.

Panels (c) and (f) show the model uncertainty in the MECP singlet-triplet gap and the activation energy, respectively. Both are quite low, indicating high model confidence for the derivatives studied here. The mean uncertainties are 0.21 kcal/mol and 0.41 kcal/mol for the singlet-triplet gap and activation energy, respectively. This is consistent with the error trends in Table \ref{tab:model_accuracy}. The uncertainty should be interpreted with caution, however, as neural network ensembles tend to be overconfident \cite{kahle2022quality}. Large uncertainty necessarily means high error, as demonstrated by the error reduction in Table \ref{tab:model_accuracy} when excluding the most uncertain geometries; however, low uncertainty does not guarantee low error. A more detailed examination of uncertainty, including calibration to the observed error and use of different architectures in the ensemble, is left to future work.

\subsubsection*{Targeting desired properties}
Absorption wavelength and thermal stability are two key properties in the design of photoactive drugs. The preferred absorption range is generally 650 to 900 nm, since human tissue is transparent only in this narrow region of the near-IR \cite{lerch2016emerging}. For photoactive drugs one typically wants the isomerization barrier to be as high as possible, so that the unstable isomer is active for as long as possible. For ion channel blockers, by contrast, the target lifetime is usually milliseconds \cite{mourot2011tuning}. For reference, the half-life of azobenzene is 1.4 days in benzene solution at $35^{\circ}$C \cite{talaty1967thermal}. Here we use the screening results to identify redshifted derivatives with high or low barriers.

Figure \ref{fig:lambda_and_barrier} shows combinations of $\lambda_{\mathrm{cis}}$, $\lambda_{\mathrm{trans}}$, and $\Delta G^{\mathrm{eff}}$, where $\lambda_i$ is the absorption wavelength of isomer $i$. We found that \textit{trans} is the more stable isomer for 99.6\% of all species. The stable isomer is usually the one activated by light, and so $\lambda_{\mathrm{trans}}$ is usually the quantity of interest. Panels (a) and (c) show that redshifting to the near-IR is quite difficult. Out of 19,000 compounds, only five have $\lambda_{\mathrm{trans}} > 600$ nm. Of these, only one has a barrier greater than that of azobenzene. There are 1,641 species with $\lambda_{\mathrm{trans}} > 500$ nm (8.7\%), including 475  with a barrier greater than azobenzene (2.5\%). However, the majority of predictions over 500 nm are significant overestimates. As discussed below, most are actually closer to 470 nm.

Two redshifted species are shown in panels (d) and (e). Compound \textbf{7} has a high barrier and compound \textbf{8} has a low barrier (top and bottom, respectively). We confirmed these predictions using single-point SF-TDDFT calculations. The quantum chemical activation free energies are shown below the model predictions, and the two agree quite well. We used the model results for the quasi-harmonic and conformational contributions to the enthalpy and entropy, since a numerical Hessian with SF-TDDFT would be prohibitively expensive.

While the barriers agree quite well with SF-TDDFT, the \textit{trans} absorption wavelengths are overestimated. For example, the predicted and true wavelengths are 567 nm and 467 nm for compound \textbf{7}, and 574 and 533 nm for compound \textbf{8}. Moreover, this latter result is somewhat suspect, since the first excited state with SF-TDDFT has square spin $\langle S^2 \rangle = 1.2$, indicating high spin contamination.  Indeed, a restricted TDDFT calculation with the $\omega$B97X-D3 functional \cite{chai2008long} and the def2-SVP basis \cite{weigend2005balanced} yielded $\lambda_{\mathrm{trans}} = 470$ nm. These results are common: after performing calculations for the 40 species with the highest \textit{trans} absorption wavelengths, we found that all had either true values around 470 nm, or had significant spin contamination leading to untrustworthy results.

We note that while the $S_1$ spin contamination was severe for some species with ultra-high absorption wavelengths, it was otherwise low in general. Indeed, the average square spin of the $S_1$ state was 0.37 in the training set. This is higher than the mean value of 0.16 for the $S_0$ state, but still reasonable. The maximum $\langle S^2 \rangle$ allowed in the training set was 1.5 for the $S_1$ state. To avoid any $S_1$ spin contamination, one could always fine-tune the model with a small dataset of excitation energies from spin-complete TDDFT and $\omega$B97X-D3. Multi-reference effects would likely be small for equilibrium structures, and only a few new calculations would be needed for fine-tuning \cite{ang2021active, axelrod2022excited}.

These results lead to several conclusions. First, redshifting is quite difficult. This is shown by the fact that most model predictions above 500 nm are actually error outliers. The associated wavelengths are still much higher than the base compound, but quite far from the predicted values. Second, we find that spin contamination is severe for many of the high-$\lambda$ predictions. Third, on a positive note, the model is able to identify redshifted species with either high or low barriers, even though the redshift is overestimated. This is encouraging for virtual screening in photopharmacology.

The quantum chemistry absorption wavelengths also come with several sources of uncertainty. They include errors in SF-TDDFT, implicit treatment of the solvent, and use of static structures instead of thermally sampled geometries \cite{axelrod2022excited}. Moreover, since the experimental absorption width is usually quite large, compounds can often absorb at wavelengths 100-200 nm higher than their peak \cite{dong2017near}.

\subsubsection*{Graph-property relationships}


Here we analyze the relationship between substitution and chemical properties. Observing and explaining general trends will enable more focused candidate screening in the future. Previous papers have also explored these relationships computationally \cite{dokic2009quantum, knie2014ortho, schweighauser2015attraction, liu2018theoretical, mukadum2021efficient}; we build on their conclusions and extend them to other substitution patterns and groups.

Figure \ref{fig:results_by_pattern} shows barrier heights and absorption wavelengths by substitution pattern. Panel (a) shows that motifs A, F, and G have bimodal $\Delta G^{\mathrm{eff}}$ distributions, with high barriers around the second mode. This is explored more below. Patterns B, C,  and E have below-average barriers and elongated distributions, while pattern H has a tight distribution and the lowest mean barrier.

Panels (b) and (c) show $\lambda_{\mathrm{trans}}$ and $\lambda_{\mathrm{cis}}$, respectively. Intriguingly, we see that the distributions of classes E and H  are very elongated for $\lambda_{\mathrm{trans}}$. The same is true to a lesser extent for B and C. While patterns like A, D, and F are tightly concentrated between 400 and 425 nm, class E samples from 400 to 525 nm with almost equal probability. The $\lambda_{\mathrm{cis}}$ distributions are much tighter.

To better understand these results, we next analyze the relationship between substituent properties and molecule properties for pattern A. Figure \ref{fig:barrier_trends_A}(a) shows that $\lambda_{\mathrm{trans}}$ is maximized when $R_1$ is a non-ring donor. In particular, panel (b) shows that this occurs when both $R_1$ and $R_2$ are strong donors. Each box in this panel shows the root-mean-square of $\lambda_{\mathrm{trans}}$ for the given ($R_1$, $R_2$) pair, computed as $\mathrm{min} \{ \lambda_{\mathrm{trans}}  \} + \mathrm{mean} \{ (\lambda_{\mathrm{trans}} - \mathrm{min} \{ \lambda_{\mathrm{trans}}  \})^2 \}^{0.5}$. This gives a mean that is weighted towards higher values, reflecting our interest in maximizing $\lambda_{\mathrm{trans}}$ better than a simple mean. The results are somewhat unexpected: a simple picture of donors raising the HOMO and acceptors lowering the LUMO would predict a redshift for strong donors \textit{or} acceptors. Yet the actual effect is only noticeable for donors.

Panel (c) shows that strong $R_1$ acceptors lead to very high barriers, with narrow distributions centered around 33 kcal/mol. This is consistent with Ref. \cite{knie2014ortho}, which reported long thermal lifetimes for ortho-fluoro substituted azobenzenes. Donor-substituted azobenzenes have noticeably lower barriers. However, the associated barriers are tightly concentrated near 30 kcal/mol, which is higher than that of unsubstituted azobenzene. This is encouraging, since it means that substitution with two donors can redshift the absorption wavelength without decreasing the barrier.

Panel (d) reinforces that $\Delta G^{\mathrm{eff}}$ has the opposite trend of $\lambda_{\mathrm{trans}}$ (in this plot we use the mean for each box). Again, however, we see that even the lowest barriers in the upper right corner are similar to that of azobenzene. Intriguingly, panel (e) shows the opposite trend for $\Delta G^{\dagger}$. This may be related to the absence of an inversion mechanism for the ISC approach. This result reinforces the need for high-accuracy quantum chemistry to accurately compare $\Delta G^{\dagger}$ and $\Delta G^{\mathrm{eff}}$.

Panel (f) shows the relationship between the activation free energy and the reaction free energy $\Delta G_{\mathrm{rxn}}$, where $\Delta G_{\mathrm{rxn}} = G_{\mathrm{cis}} - G_{\mathrm{trans}}$. The Bell-Evans-Polanyi principle \cite{bell1936theory, evans1937introduction} states that the reaction enthalpy and activation enthalpy are linearly related for reactions in the same family. This relationship was also tested for azobenzene derivatives in Ref. \cite{dokic2009quantum}. We see a moderate negative correlation between the two quantities, with Spearman $\rho$ equal to $-0.27$. Hence $\Delta G^{\mathrm{eff}}$ can be increased by making the \textit{cis} isomer more stable. However, the modest correlation means that this is not the full story, and that the TS and MECP energies must be explicitly considered.

\section*{Discussion}


The main source of error in our approach is the underlying quantum chemistry calculations. As discussed in SI Sec. \ref{sec:benchmark}, expensive wavefunction methods give lower barriers than  SF-TDDFT, but still give different answers from each other. On balance the most accurate method seems to be SF-EOM-CCSD(dT), but its prohibitive $N^7$ scaling \cite{manohar2008noniterative} makes it a poor candidate for transfer learning. Future work should focus on accurate quantum chemical approaches that do not need manual setup, such as selection of active spaces, and that are affordable enough for transfer learning.

Another limitation is that we have not considered azobenzene protonation and azo-hydrazone tautomerism. These effects can be facilitated by substituents such as $\mathrm{NH_2}$ and $\mathrm{OH}$, and by solvation in a protic solvent, weakening the N=N double bond and lowering the isomerization barrier \cite{kojima2005effect, matazo2008azo, rickhoff2022reversible}. Protonation from the solvent is not accounted for in a PCM description. Incorporating automated protonation tools \cite{pracht2017automated} into our workflow would be of interest in the future. Similarly, the protein environment for a given target in photopharmacology can also affect the isomerization rate \cite{gaspari2017structural, liang2022multiscale}. Incorporating these effects for a specific target, as in Ref. \cite{liang2022multiscale}, is of interest for future work.

From the perspective of property optimization, the biggest remaining challenge is redshifting. While it is straightforward to reach $\lambda = 470$ nm for \textit{trans} isomers, it appears very difficult to reach $\lambda =$ 550-600 nm. Averaging the gap over thermally sampled geometries may increase the wavelength and improve prediction accuracy \cite{axelrod2022excited}. Including bulky groups in all four ortho positions may also increase the wavelength \cite{axelrod2022excited}, but at the potential cost of synthetic accessibility. A more targeted approach to wavelength optimization could be of interest in the future. For example, methods such as Monte Carlo tree search \cite{dieb2019monte} could likely improve over virtual screening of combinatorial libraries.

\section*{Conclusions}
We have presented a fast and automated method for predicting the isomerization barriers of azobenzene derivatives. The approach can compute the activation free energy through TS theory or ISC theory. We have demonstrated the accuracy of the underlying ML model with respect to SF-TDDFT, reproduced trends in the experimental isomerization rate, and argued for rotation-based ISC as the reaction mechanism. Our software is fast, accurate, and easily accessible to the community, making it a valuable tool for computational design of photoactive molecules. Future work will focus on more accurate quantum chemistry methods and more targeted molecular generation.

\section*{Supporting Information}
Extended methods, details of molecule generation, details of optimization, non-adiabatic transition state theory, activation free energies by mechanism, thermal isomerization rates, details of training, filtering protocol in screening, collection of experimental data, note on mechanisms, and quantum chemistry benchmark.

\section*{Acknowledgements} \label{sec:acknowledgements}
We thank Dr. Johannes Dietschreit for bringing triplet-mediated isomerization to our attention. Harvard Cannon cluster, MIT Engaging cluster, and MIT Lincoln Lab Supercloud cluster at MGHPCC are gratefully acknowledged for computational
resources and support. Financial support from DARPA (Award HR00111920025) is acknowledged.


\clearpage
\newpage
\onecolumn


 

\setcounter{page}{1}
\renewcommand{\thepage}{S\arabic{page}}

\renewcommand\thesection{S\arabic{section}}
\renewcommand\thesubsubsection{\Alph{subsubsection}}
\renewcommand\theequation{S\arabic{equation}}
\setcounter{section}{0} 
\setcounter{subsection}{0} 
\setcounter{subsubsection}{0} 
\setcounter{equation}{0} 
\setcounter{table}{0}
\setcounter{figure}{0}
\renewcommand\thefigure{S\arabic{figure}}
\renewcommand\thetable{S\arabic{table}}

\onecolumn
\begin{center}
\section*{Supporting Information} 
\section*{Thermal half-lives of azobenzene derivatives: virtual screening based on intersystem crossing modeled with a machine learning potential }
\end{center}

\section{Extended methods}
\label{sec:si_ex_methods}

All models used the PaiNN architecture \cite{schutt2021equivariant} and were implemented in PyTorch \cite{NEURIPS2019_9015}. As in our previous work, we used five convolutions instead of the three used originally, as this substantially improved model performance \cite{axelrod2022excited}. We also allowed the $k$ values in the radial basis functions to be updated during training. Remaining hyperparameters can be found in Ref. \cite{axelrod2022excited}, and an in-depth explanation of their meaning can be found in Ref. \cite{schutt2021equivariant}. Further hyperparameter optimization is likely possible \cite{frey2022neural}.

After training the model on data without empirical dispersion, we added D3BJ dispersion \cite{grimme2010consistent, grimme2011effect} to the total energy prediction:
\begin{align}
    & E = E_{\mathrm{model}} + E_{\mathrm{D3}} \\ 
    & \mathbf{F} = -\nabla E_{\mathrm{model}} - \nabla E_{\mathrm{D3}}
\end{align}
The forces were computed with automatic differentation \cite{paszke2017automatic}. Dispersion parameters were taken from the PhysNet repository \cite{unke2019physnet, physnet_git}. No cutoff was used. D3 dispersion is a function only of the atom types and positions, and could therefore be added analytically without approximation. The model performance in the main text used SF-D3BJ TDDFT as the ground truth.

The geometries for SF-TDDFT/C-PCM calculations were generated through active learning (Fig. \ref{fig:ts_and_al}(a)). In each round of active learning, the TS workflow in Fig. \ref{fig:ts_and_al}(b) was performed for 1,000 (SMILES, mechanism) pairs using one of the trained models. The pairs were randomly sampled from a set of 73,000 reactions with four mechanisms each. We selected 3,000 generated geometries for DFT calculations, added the new data to the training set, and repeated. 50\% of the geometries were selected by uncertainty, 30\% by energy, and 20\% randomly. For uncertainty-based selection, we chose the geometries with the highest variance in forces predicted by the three models. For energy selection, we sampled a geometry $i$ with probability $p_i \propto e^{E_i / k_{\mathrm{B}} T}$. The energy was computed relative to other geometries of the same species in the same simulation. The energy selection sampled near-TS geometries from relaxed scans, and high-energy MTD structures from conformer generation.

Active learning was repeated 15 times, yielding 43,938 calculations in total. The final model was trained on 42,938 geometries, with 500 used for validation and 500 for testing.

\section{Molecule generation}
\label{sm_sec:molgen}
Molecules were generated using the patterns in Fig. \ref{fig:results_by_pattern}(d). These are common substitution patterns in the literature; see, for example, the synthesized species in Refs. \cite{malkin1962temperature, nagamani2005photoinduced, sierocki2006photoisomerization, dokic2009quantum, bandara2010proof,  bandara2011short, beharry2011azobenzene, knie2014ortho, schweighauser2015attraction, moreno2016sensitized}. Some of the functional groups were common azobenzene substituents aggregated in Ref. \cite{axelrod2022excited}, with new groups added from Refs. \cite{gutzeit2021fine, lv2021computational, konrad2020computational}. Others were basic chemical moieties, such as benzene, ethane, $\mathrm{\mathrm{R}-CF_3} $, and so on. 

We aimed to avoid steric clashes in tetra-ortho substitution, since this can make experimental synthesis more difficult. For this reason we separated substituents by size. Those with four total atoms or fewer were labeled small, and all others were labeled large. For all tetra-ortho substitutions, we only allowed one side of the N=N bond to be substituted with large groups. The other side could only be substituted with small groups. Substitution with small groups at all positions was also allowed.

\section{Optimizations}
\subsection{Relaxed scans}
To perform the relaxed scans, we first identified the CNNC atoms using a substructure search of azobenzene in RDKit \cite{rdkit}. We then adjusted their angles and/or dihedral angle by a constant amount in each step of the scan. The angles and dihedrals were adjusted using  the atomic simulation environment (ASE) \cite{ase-paper}. For the inversion mechanism we set the final CNN or NNC angle to 179.5$^{\circ}$. We used 179.5$^{\circ}$ instead of 180$^{\circ}$ because the Cartesian derivative of the latter is undefined, and hence the constraining forces described below would also be undefined. For rotation we set the final dihedral angle to $\pm 90^{\circ}$. We also set the final CNN and NNC angles to 122$^{\circ}$. Without this constraint, we found that some of the scans collapsed to inversion TSs instead of rotation TSs. 122$^{\circ}$ is the angle of the rotational TS when optimizing azobenzene with SF-TDDFT \cite{yue2018performance}.

20 steps were used in each scan. After the angles and/or dihedrals were adjusted, the geometry was optimized with forces given by $\mathbf{F} + \mathbf{F}_{\mathrm{restrain}}$. The restraining forces are the negative gradients of the following restraining energies:
\begin{align}
    & E_{\alpha, n} = k_{\alpha} (\alpha - \alpha_n)^2 \\
    & E_{\omega, n} =  -k_{\omega} \mathspace \mathrm{cos} \mathspace (\omega - \omega_n).
\end{align}
Here $\alpha$ denotes an angle, $\omega$ denotes a dihedral, $n$ denotes the $n^{\mathrm{th}}$ step of the optimization, and $\alpha_n, \ \omega_n$ are target values at the $n^{\mathrm{th}}$ step. $E_{\alpha}$ was used for the inversion mechanism, while both $E_{\alpha}$ and $E_{\omega}$ were used for the rotation mechanism.

$k_{\alpha}$ and $k_{\omega}$ were each set to 1 Ha. The optimization was performed with the BFGS algorithm \cite{broyden1970convergence, fletcher1970new, goldfarb1970family, shanno1970conditioning} in ASE. A rather loose convergence threshold of $f_{\mathrm{max}} = 0.05$ eV/\AA \ was used. This was because of the conformer generation performed after the scans. As described in Sec. \ref{sec:confgen}, each stage of conformer generation was restarted if a new conformer had a lower energy than the seed conformer. We used a tolerance of $f_{\mathrm{max}} = 0.05$ eV/\AA \ to minimize the optimization time spent on high-energy conformers. Tight convergence criteria were only used once no lower energy conformers were found. If the seed conformer were optimized with tighter thresholds than the generated ones, it could have a lower energy simply due to its stricter thresholds. Hence it needed to be optimized with the same thresholds.

Angles and dihedrals were set with ASE at each step. ASE adjusts only these internal coordinates, without rotating the rest of the molecule as a solid body. RDKit uses a solid body rotation, and in principle should then require fewer optimization steps at each stage of the scan. However, we found that this led to unstable scans, and so we used ASE instead.

For all optimizations and dynamic simulations we updated the neighbor list every ten steps. To account for neighbors coming into the cutoff radius between updates, we computed the neighbors using a cutoff of $r_{\mathrm{cut}} + r_{\mathrm{skin}}$, where $r_{\mathrm{cut}}=5.0$ \AA \ is the model cutoff radius and $r_{\mathrm{skin}} = 2.0$ \AA \ is the cutoff skin. Distances between atoms and neighbors were computed at each step, and only those within $r_{\mathrm{cut}}$ of each other were used in the model.

\subsection{Conformer generation}
\label{sec:confgen}

\begin{figure*}[t!]
    \centering
    \includegraphics[width=\textwidth]{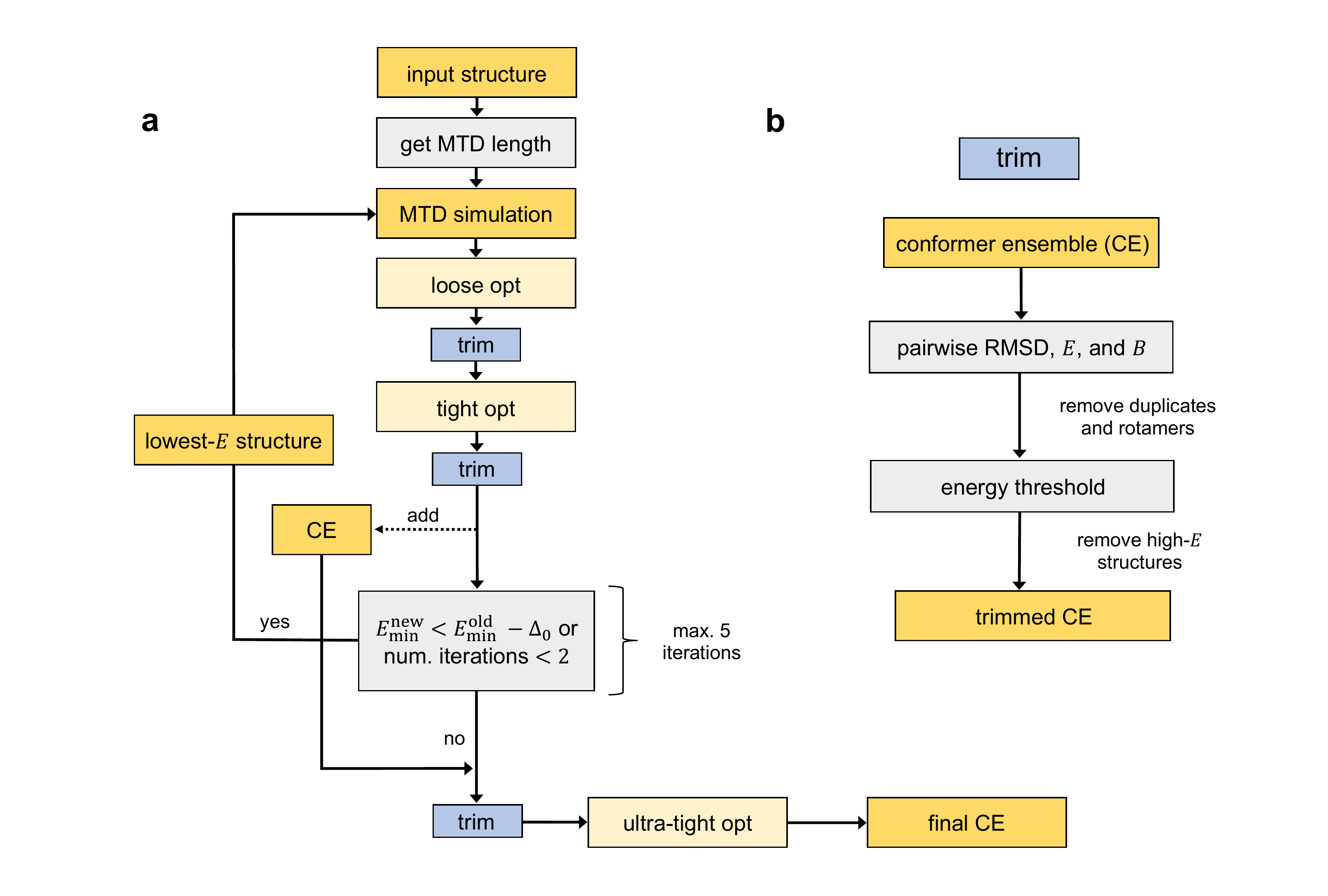}
    \caption{Outline of the conformer generation workflow used in this work. (a) Overview of the approach. (b) Details of the trimming procedure. }
    \label{fig:conformer_search}
\end{figure*}

\begin{table*}[t!]
\small
\centering
\begin{tabular}{c|c|c}
     \hline
     Parameter & Meaning & Value \\
     \hline
     $t_{\mathrm{new}}$ & time between adding reference structures to $V_{\mathrm{bias}}$ & 1 ps \\
     $\kappa$ & turn-on parameter for adding new MTD Gaussians \cite{grimme2019exploration} & 0.03 \\
     $k'$ & $k= N k'$ is the biasing strength in Eq. (\ref{eq:v_bias}) \cite{pracht2020automated} &  1.5 m$E_{\mathrm{h}}$ \\
     $\alpha$ & Gaussian width in Eq. (\ref{eq:v_bias}) & 0.5 $\mathrm{Bohr}^{-2}$ \\ 
     $\Delta t$ & time step & 2 fs \\
     $m_{\mathrm{H}}$ & hydrogen mass used in dynamics & 4 amu \\
     thermostat &  method for enforcing average temperature during dynamics & Nosé-Hoover \\
     $Q$ & Nosé-Hoover effective mass & $(3 N - 6) \cdot \tau^2  k_{\mathrm{B}} T $ \\
     $\tau$ & Nosé-Hoover relaxation time & 100 fs  \\
     $T$ & simulation temperature & $298.15 \ \mathrm{K}$  \\
     \hline
     $\Delta_0$ & see Fig. \ref{fig:conformer_search}(a) & 0.2 kcal/mol \\
     $f_{\mathrm{max}}^{\mathrm{loose}}$ & maximum force component in loose opt &  0.2 eV/\AA \\
     $f_{\mathrm{max}}^{\mathrm{tight}}$ & maximum force component in tight opt &  0.05 eV/\AA \\
     $f_{\mathrm{max}}^{\mathrm{ultra}}$ & maximum force component in ultra-tight opt & 0.005 eV/\AA \\
     $E_{\mathrm{win}}^{\mathrm{loose}}$ & energy window for retaining conformers after loose opt & 15.0 $\mathrm{kcal} / \mathrm{mol}$ \\
     $E_{\mathrm{win}}^{\mathrm{tight}}$ & energy window for retaining conformers after tight opt & 8.0 $\mathrm{kcal} / \mathrm{mol}$ \\
     $E_{\mathrm{win}}^{\mathrm{ultra}}$ & energy window for retaining conformers after ultra-tight opt & 2.5 $\mathrm{kcal} / \mathrm{mol}$ \\
     \hline 
     $\Delta E_{\mathrm{thr}}$ & two geometries cannot be duplicates or rotamers if their energies differ by more than $\Delta E_{\mathrm{thr}}$ & 0.15 kcal/mol \\
     $\Delta R_{\mathrm{thr}}$ & two geometries cannot be duplicates if their RMSD is more than $\Delta R_{\mathrm{thr}}$ & 0.175 \AA \\
     $\Delta B_{\mathrm{thr}}$ & two geometries cannot be rotamers if their rotational constants differ by more than $\Delta B_{\mathrm{thr}}$ & 0.03 (3\%) \\
    \hline
\end{tabular}
\caption{Parameters used in our NN conformer generation workflow. From top to bottom, the sections contain MTD parameters, optimization parameters, and \texttt{cregen} parameters. $N$ is the number of atoms in the system. The \texttt{cregen} parameters are the defaults used in CENSO, a program that refines CREST ensembles with DFT \cite{grimme2021efficient}.}
\label{tab:confgen_params}
\end{table*}

Conformer searches were performed for all reactants, products, and transition states. We implemented a method based on CREST \cite{pracht2020automated} using our NN. CREST combines metadynamics, high-temperature molecular dynamics, and genetic structure crossing \cite{grimme2017fully} to sample conformational space. The sampled geometries are then optimized with a multi-level filtering scheme. In this approach, optimizations are performed with progressively tighter thresholds, and structures are discarded if their energy is above a maximum value that is lowered in each step. By default CREST uses the fast semi-empirical method GFN2-xTB \cite{bannwarth2019gfn2} to compute energies and forces.

The key component for phase space exploration is metadynamics (MTD). The collective variables used for MTD are the root-mean-square displacements (RMSDs) from previously visited structures. This drives the molecule away from previously visited configurations and towards new regions of phase space. The associated biasing potential is given by
\begin{align}
V_{\mathrm{bias}} = \sum_{i}^{n} k_{i} \mathspace \mathrm{exp}(-\alpha_i \Delta_i^2). \label{eq:v_bias}
\end{align}
Here the sum is over $n$ previously visited structures, where a new structure is added every 1 ps. $\Delta_i$ is the RMSD of the current structure with respect to the $i^{\mathrm{th}}$ previous structure, $k_i$ is a pushing strength parameter, and $\alpha_i$ is a width parameter. The forces at each step are the sum of the actual forces and the negative gradient of $V_{\mathrm{bias}}$. 14 MTD runs are performed in CREST, using different combinations of $\alpha_i$ and $k_i$.

Initially we used CREST to perform conformer searches. However, given the speed of relaxed scans and eigenvector following with our NN, we found that CREST was by far the most time-consuming step in the workflow. For example, with access to 24 Xeon-P8 CPU nodes and 1,152 cores in total, we ran 82 CREST jobs at a time with 14 cores per job. The average job took 4.2 hours, which meant a throughput of 470 geometries per day. Given four TS guesses per species, plus the reactant and product geometries, this meant that only 78 species could be screened per day. Screening 25,000 species, as done in this work, would have taken 11 months.

CREST has built-in options for conformer searches that are faster but less extensive. These involve fewer MTD runs and tighter energy windows for each level of optimization. However, given less extensive sampling and tighter energy windows, it makes sense to use our NN instead xTB. For example, CREST has a default energy window of 6.0 kcal/mol for the final ensemble, even though conformers of energy greater than approximately $2.5$ kcal/mol are not populated at room temperature. This is to compensate for errors in xTB, so that low-lying conformers mistakenly assigned a high energy by xTB are not discarded. These conformers can be subsequently re-ranked or re-optimized with a higher level of theory. Therefore, it is more reasonable to use a tight energy window for the NN, which is specifically trained to reproduce SF-TDDFT for azobenzene derivatives. Similarly, the sampling should be more extensive for xTB than for the NN. This is because errors in xTB mean that conformational space must be thoroughly sampled up to an energy of 6.0 kcal/mol. The NN, by contrast, only needs to thoroughly sample the space up to an energy of 2.5 kcal/mol. Alternatively, one could retain the extensive sampling but use the ultra-fast GFN force-field \cite{spicher2020robust}. However, we found that the force-field produced poor results for TSs.

We therefore implemented our own conformer search with the NN, and used tighter energy windows and fewer dynamics simulations. The workflow is shown schematically in Fig. \ref{fig:conformer_search}. The approach closely follows that of CREST, but uses only one MTD run per stage, and does not use MD or genetic structure crossing. As in CREST, we determined the MTD time using a flexibility measure derived from the chemical graph (the current formula in CREST differs from that of the original publication; we used the up-to-date version in the source code \cite{crest_git}, commit 6bb6355). To partially compensate for the decreased total simulation time, we used a minimum MTD time of 15 ps, instead of the 5 ps used in CREST. The parameters used in our workflow are given in Table \ref{tab:confgen_params}.

Several points require further discussion. First, we used the CREST \texttt{cregen} tool to remove both duplicates and rotamers in the ensemble. Duplicates are pairs of geometries with $\Delta E < \Delta E_{\mathrm{thr}}$, $\mathrm{RMSD} < R_{\mathrm{thr}}$, and $B < B_{\mathrm{thr}}$, where $\Delta$ is the difference between the two quantities, $E$ is the energy, $B$ is the rotational constant, and $\mathrm{thr}$ denotes a threshold. Rotamers are structures with $\Delta E < \Delta E_{\mathrm{thr}}$, $\mathrm{RMSD} > R_{\mathrm{thr}}$, and $B < B_{\mathrm{thr}}$. We removed rotamers because the difference in rotamer count per conformer should be small, and because accounting for all rotamers is quite difficult. This is especially true when MD and genetic structure crossing are not included, since they are included in CREST primarily to find rotamers \cite{pracht2020automated}. 

Second, we used the L-BFGS algorithm \cite{liu1989limited} in ASE to perform optimizations. The convergence criterion is $ \max_{i, \alpha} \vert f_{i, \alpha} \vert < f_{\mathrm{max}}$, where $f_{i, \alpha}$ is a force component, $i \in \{1, N \}$ is the atom index, $\alpha \in \{ x, y, z\} $ is the Cartesian index, and $f_{\mathrm{max}}$ is a threshold. For equilibrium geometries we used $f_{\mathrm{max}}^{\mathrm{ultra}} = 0.005$ eV/\AA. For TS geometries we used $f_{\mathrm{max}}^{\mathrm{ultra}} = 0.01$ eV/\AA, since the average change in energy between the two thresholds is only 0.1 kcal/mol, while the number of extra steps can be significant. We used $f_{\mathrm{max}} = 0.005$ eV/\AA \ for eigenvector following performed on the five lowest-energy conformers. Note also that the algorithm is implemented in Cartesian coordinates. This makes it less efficient than the internal optimizer in xTB, which uses internal coordinates. Use of an internal coordinate optimizer may be of interest in the future.

Third, unlike in CREST \cite{grimme2019exploration}, we did not constrain bond lengths with SHAKE \cite{ryckaert1977numerical}.  Typically SHAKE is used to allow longer time steps and therefore accelerate the dynamics. Indeed, CREST uses a default time step of 5 fs, while the maximum value for unconstrained dynamics is typically 0.5 fs. However, the SHAKE implementation in ASE is rather slow, and therefore became a bottleneck instead of reducing run times. One can also increase the time step by artificially increasing the mass of hydrogen, since, as the lightest element, its motion is usually the fastest in the system. We followed the default in CREST and set the hydrogen mass to 4 amu. This allowed us to use a time step of 2 fs. Interestingly, we found that CREST automatically reduced the time step from 5 fs to 2 fs for TS conformer searches, since trial MTD runs with larger time steps all failed. This happened even though SHAKE is used in CREST. The same thing did not happen for conformer searches of equilibrium geometries. Hence our time step was the same as the CREST time step for TSs, even though we did not use SHAKE. Yet an efficient implementation of SHAKE is still of interest, since fixing the bond lengths during MTD might decrease the number of subsequent optimization steps needed.

Fourth, we fixed the CNNC atoms in each molecule when performing TS conformer searches. We experimented with Hookean constraints on the CNN angles and CNNC dihedrals, but this led to unstable dynamics. This was likely because the Cartesian gradient of an angle is undefined when the angle is $180^{\circ}$. We excluded the fixed atoms from the RMSD computation in MTD. For equilibrium geometries we excluded the CNNC atoms from the RMSD, but did not fix the atoms. Excluding these atoms ensured that $\textit{cis}$ did not isomerize to $\textit{trans}$.

\begin{figure*}[t]
    \centering
    \includegraphics[width=\textwidth]{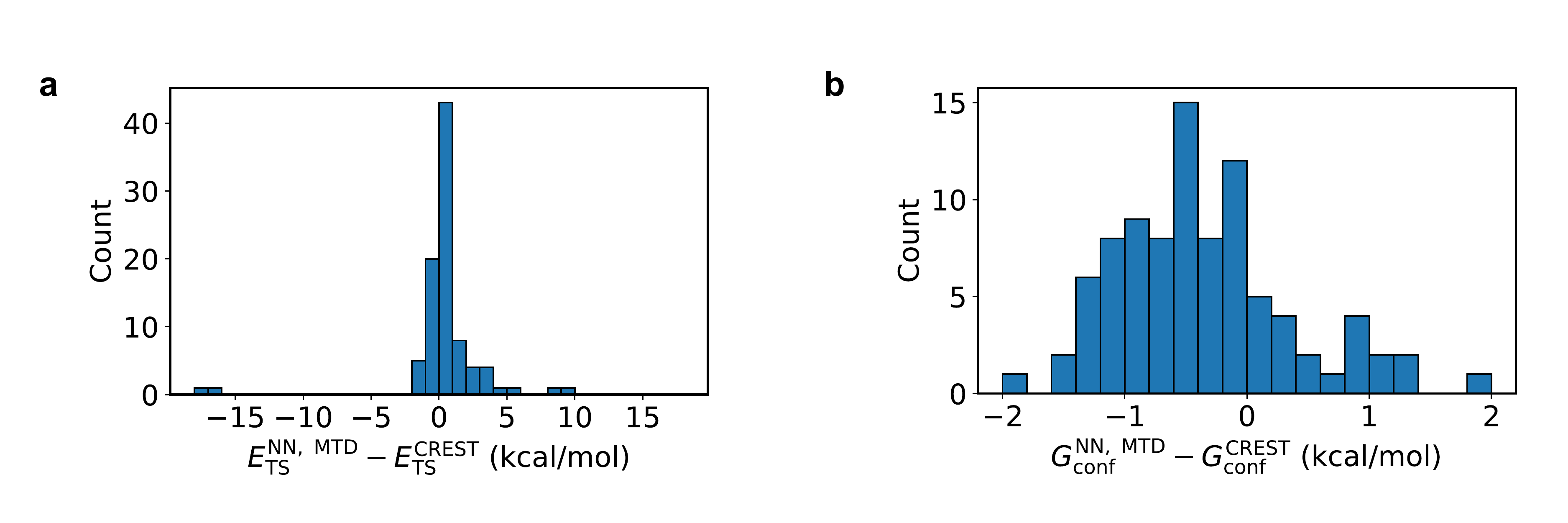}
    \caption{Energy differences between TSs generated with NN conformer searches and with CREST. }
    \label{fig:conf_accuracy}
\end{figure*}

To evaluate our approach, we performed conformer searches with CREST and with our method on 100 relaxed scan TS geometries. We then optimized each CREST geometry with the NN, removed duplicates, and retained the geometries with energies under 2.5 kcal/mol. The top 5 conformers from each method were then optimized with eigenvector following. We checked that the mechanism of each TS matched the target mechanism of the relaxed scan (e.g. rotational TS for a rotation relaxed scan). We removed one species that had relaxed to an inversion TS with CREST after being seeded with a rotational TS. The distribution of TS energy differences is shown in Fig. \ref{fig:conf_accuracy}(a). The mean value of $E_{\mathrm{TS}}^{\mathrm{NN, \ MTD}} - E_{\mathrm{TS}}^{\mathrm{CREST}}$ is 0.28 kcal/mol. The mean increases to 0.48 kcal/mol after removing outliers with absolute energy differences over 6 kcal/mol. The agreement between the two methods is quite good, despite using only one MTD simulation in the NN approach instead of 14. Further, the error in the $\Delta E^{\dagger}$ is expected to be smaller than the error in $E_{\mathrm{TS}}$. This is because the conformer search can only overestimate the lowest energy, not underestimate it, and so some error cancellation can be expected.

We also compared the completeness of the ensembles in each of the two methods. To do so we computed the conformational free energy, defined in SI Sec. \ref{sm:free_energy}. Figure \ref{fig:conf_accuracy}(b) shows the difference in conformational free energy between the two methods. Interestingly, we see that the NN generates a more complete ensemble on average. The mean value of $G_{\mathrm{conf}}^{\mathrm{NN, \ MTD}} - G_{\mathrm{conf}}^{\mathrm{CREST}} $ is $-0.36$ kcal/mol. This trend persists even when the lowest-energy conformers are quite close in energy between the two methods. For example, restricting ourselves to species in which the $\vert E_{\mathrm{TS}}^{\mathrm{NN, \ MTD}} - E_{\mathrm{TS}}^{\mathrm{CREST}} \vert \leq 0.5 $ kcal/mol, we find that the difference in $G_{\mathrm{conf}}$ is $-0.60$ kcal/mol. We conclude that the NN ensembles are quite complete, and that the associated $G_{\mathrm{conf}}$ is reliable.

The NN approach is faster than CREST, mainly because fewer MTD runs are performed. NN energy and force calculations also tend to be faster than xTB for large molecules. A precise comparison is difficult, because xTB is run on CPUs and NNs are run on GPUs. This means that one cannot simply compare wall-clock times using equivalent hardware. Indeed, to compare the number of species that can be screened per day, one must also consider the number of GPUs available to the user vs. the number of CPUs. Nevertheless, some rough comparisons can still be made. For example, running 189 ps of MTD for a molecule with 102 atoms took 24 hours and 48 minutes with xTB on one Xeon-P8 core. If the xTB calculations were parallelized over all 48 cores on the node, then the run would take 31 minutes given perfect parallelization. Hence 1.9 MTD runs could be performed per node per hour. By contrast, the corresponding NN MTD run took 4 hours with one NVIDIA A100 GPU, using a batch of 50 species. This corresponds to 12.5 MTD runs per GPU per hour, or 25 per node per hour, assuming 2 GPUs per node. In this case, NN MTD is effectively $ 25 / 1.9 \approx 13$ times faster than xTB MTD.

Lastly, since the conformer search is the rate-limiting step in our workflow, any additional speed-ups would increase our throughput. While MTD enables exploration of conformational space, it is fundamentally limited by the fact that all steps must be taken in series. Uncorrelated geometries can only be generated after taking several hundred steps. Stochastic conformer generators are attractive because they can rapidly produce many uncorrelated structures. Examples include the ETKDG method in RDKit \cite{riniker2015better} and the commercial software Omega \cite{hawkins2010conformer, hawkins2012conformer}. To adapt these generators for TSs, one would have to generate conformers of equilibrium geometries, and perform a relaxed scan from the equilibrium structures to the TS. Even if TS geometries would be generated directly, the structures would still have to be optimized. In either case, optimization would incur significant cost. Indeed, structure optimizations take the most time of any step in CREST and NN conformer generation. Therefore, avoiding MTD would not reduce the computation time to that of a stochastic conformer generator. Another approach is to train a generative model to produce TS geometries \cite{pattanaik2020generating}. ML for equilibrium conformer generation has rapidly progressed in recent years \cite{xu2020learning, shi2021learning, ganea2021geomol, xu2021end, roney2021generating, luo2021predicting, jing2022torsional, xu2022geodiff, hoogeboom2022equivariant}, and many of the methods could also be applied to TS generation. These avenues would certainly be of interest in future work.

\subsection{Eigenvector following}
EVF was implemented with Baker's rational function optimization method \cite{baker1986algorithm}. In the first step, the numerical Hessian was computed with finite differences, using the ASE vibrations package. We used a step size of 0.005 \AA, and checked that the results were indistinguishable from the analytical Hessian computed with PyTorch. We used finite differences rather than automatic differentiation to minimize the consumed GPU memory. In subsequent steps the Hessian was updated using Powell's method \cite{powell1971recent}. We used a convergence threshold of $f_{\mathrm{max}} = 0.005$ eV/\AA. 

\subsection{Intrinsic reaction coordinate}
\begin{table*}[t]
\small
\centering
\begin{tabular}{c|c|c}
     \hline
     Parameter & Meaning & Value \\
     \hline
     \texttt{init\_displ\_de} & Target energy change for the first step & 0.25 kcal/mol \\
     \texttt{scale\_displ\_sd} & Factor for scaling the first SD step & 0.15 \\
     \texttt{adapt\_scale\_disp} & Modify \texttt{scale\_displ\_sd} when the step size becomes smaller or larger & True \\
     \texttt{sd\_parabolic\_fit} & Do a parabolic fit for finding the optimal SD step length & True \\
     \texttt{interpolate\_only} & Only allow interpolation for parabolic fit, not extrapolation & True \\
     \texttt{do\_sd\_corr} & Apply a correction to the first SD step & True \\
     \texttt{scale\_displ\_sd\_corr} & Factor for scaling the correction to the SD step & 0.33 \\
     \texttt{sd\_corr\_parabolic\_fit} & Do a parabolic fit for finding the optimal SD step length & True \\
     \texttt{tol\_max\_g} & Maximum gradient for convergence & $2 \times 10^{-3}$ Ha/Bohr \\
     \texttt{tol\_rms\_g} & RMS gradient for convergence & $5 \times 10^{-4}$ Ha/Bohr \\
     \hline
\end{tabular}
\caption{Parameters used in IRC searches. SD stands for steepest descent.}
\label{tab:irc_params}
\end{table*}

We implemented an IRC algorithm based on the approach in Orca \cite{neese2020orca}. This approach is itself based on the method of Morokuma and coworkers \cite{ishida1977intrinsic}. Details can be found in the Orca 5.0.2 manual. The parameters that we used are given in Table \ref{tab:irc_params}. Note that the force tolerances are rather large, since only a loose optimization is needed to see if the IRC has the product and reactant on either end.

\subsection{Singlet-triplet crossing searches}
\label{sm_sec:mecp}
Singlet-triplet crossings were located with a two-step procedure. In the first step, we performed IRC to locate the crossings one either side of the TS. In the second step, we optimized each of the two crossings by minimizing the energy while keeping the singlet-triplet gap close to zero. This yielded MECPs on each side of the TS.

In the first step we performed the normal IRC algorithm on the $S_0$ state. We also kept track of the $T_1$ energy, and stopped the optimization once $E_{T_1} - E_{S_0}$ changed sign. We then used the final geometry as the starting point for the MECP optimization.

The MECPs were optimized with the approach of Ref \cite{lykhin2016nonadiabatic}. In a regular optimization, the objective function is the energy $E$, and so its negative gradient is $\mathbf{F}$. In an MECP optimization, the objective function is the energy subject to the constraint that the singlet-triplet gap is zero. We therefore used the following effective forces in the optimization:
\begin{align}
    \mathbf{F}' = \mathbf{P} \bar{\mathbf{F}} + \frac{1}{\alpha} \mathspace \Delta E \mathspace \Delta \mathbf{F}.  \label{eq:f_prime}
\end{align}
The first term is the mean of the singlet and triplet forces, $\bar{\mathbf{F}}$, with the gradient of the energy gap $\Delta E$ projected out by $\mathbf{P}$. Its effect is to minimize the average singlet and triplet energies without increasing their gap. The second term contains the force difference $\Delta \mathbf{F}$. Since it is scaled by the energy difference $\Delta E$, its effect is to minimize the singlet-triplet gap. $\alpha$ is a constant with units of energy. The terms are given by
\begin{align}
    & \Delta E = E_{T_1} - E_{S_0} \\ 
    & \Delta \mathbf{F} = \mathbf{F}_{T_1} - \mathbf{F}_{S_0} \label{eq:df} \\ 
    & \bar{\mathbf{F}} = \frac{1}{2} \left( \mathbf{F}_{T_1} + \mathbf{F}_{S_0} \right)  \\ 
    & \mathbf{P}\bar{\mathbf{F}} = \bar{\mathbf{F}} - \frac{\Delta \mathbf{F}^T \bar{\mathbf{F}}}{\Delta \mathbf{F}^T \Delta \mathbf{F}}  \mathspace \Delta \mathbf{F},
\end{align}
where a superscript $T$ denotes transposition. 

We implemented this approach in ASE, using a calculator that produced the forces of Eq. (\ref{eq:f_prime}). We then performed the optimization using the BFGS algorithm in ASE. The maximum force tolerance for convergence was set to 0.005 eV/\AA.  The constant $\alpha$ was set to 12.55 kcal/mol following a related MECP algorithm for conical intersections \cite{levine2008optimizing}. We found that the optimized MECPs were insensitive to the choice of $\alpha$. For example, nearly identical MECP energies were obtained when using $\alpha = 1.0 $ kcal/mol. 

\subsection{Free energy calculations}
\label{sm:free_energy}
The free energy was computed for reactants, products, and TSs, using
\begin{align}
    G = H - T S, \label{eq:g_total}
\end{align}
where $G$ is the Gibbs free energy, $H$ is the enthalpy, and $S$ is the entropy. The enthalpy was computed as
\begin{align}
	H = E_{\mathrm{el}}  + E_{\mathrm{ZPE}} + E_{\mathrm{vib}}(T) + 4 \mathspace k_{\mathrm{B}} T + E_{\mathrm{conf}}, \label{eq:enthalpy}
\end{align}
where the last term is the average conformer energy,
\begin{align}
	E_{\mathrm{conf}} = \sum_i p_i E_i,
\end{align}
and the sum is over all conformers. $p_i$ is the Boltzmann probability for conformer $i$, given by $p_i \propto \mathrm{exp}(-E_i / k_{\mathrm{B}} T)$. $E_{\mathrm{el}}$ is the electronic energy, $E_{\mathrm{ZPE}}$ is the zero-point energy, and $E_{\mathrm{vib}}(T)$ is the thermal vibrational energy. Standard expressions for $E_{\mathrm{ZPE}}$ and $E_{\mathrm{vib}}(T)$ can be found in Ref. \cite{dokic2009quantum}. Apart from $E_{\mathrm{conf}}$, all terms in Eq. (\ref{eq:enthalpy}) were computed using the ASE thermochemistry package. 

The entropy was computed as 
\begin{align}
	S = S_{\mathrm{mRRHO}} + S_{\mathrm{conf}},
	\label{eq:s_total}
\end{align}
where $S_{\mathrm{mRRHO}}$ is the modified rigid rotor-harmonic oscillator entropy (see below), and $S_{\mathrm{conf}}$ is the conformational entropy:
\begin{align}
	S_{\mathrm{conf}} = -k_{\mathrm{B}} \sum_i p_i \mathspace \mathrm{ln} \mathspace p_i. \label{eq:s_conf}
\end{align}
The sum is over all conformers of all four mechanisms when using TS theory, but only over rotational conformers when using the ISC approach. Note that if two mechanisms are the same by symmetry, then the conformational entropy is increased by $kT \mathspace  \mathrm{log} \mathspace 2$, while $\Delta G$ is lowered by this amount. This is the same factor that would arise from using the symmetry number $\sigma=2$ in the calculations  \cite{rietze2017thermal}.

The modified rigid rotor harmonic oscillator approximation \cite{grimme2012supramolecular} interpolates between a free rotor at low frequencies and a harmonic oscillator at high frequencies. This is more physically sound than a standard harmonic approximation, and avoids the divergent entropy of low-frequency harmonic modes. The rotor cutoff frequency was set to 50 $\mathrm{cm}^{-1}$, following the default in xTB \cite{bannwarth2019gfn2}. All other parameters were unchanged from Ref. \cite{grimme2012supramolecular}.

Harmonic modes and frequencies were computed in the normal way \cite{ochterski1999vibrational} by diagonalizing the mass-weighted Hessian after projecting out rotation and translation. For reactants and products we checked that there were no imaginary frequencies. For TSs we checked that there was only one imaginary frequency, with magnitude $\geq 200$ $\mathrm{cm}^{-1}$. Species with reactants or products not satisfying these conditions were removed. Species were kept if they had at least one TS geometry from each mechanism satisfying these conditions.

For $G^{\mathrm{X}}$, $H^{\mathrm{X}}$, and $S^{\mathrm{X}}$ we used thermostatistical quantities from the rotational TSs. In principle the vibrational terms should be computed with the average Hessian of the singlet and triplet states at the crossing \cite{lykhin2016nonadiabatic}. However, we found that this approach led to a sensitive dependence on $\alpha$ in Eq. (\ref{eq:f_prime}), and so we used the rotational TS results instead.

\section{Non-adiabatic transition state theory}
\label{sm_sec:na_tst}
The reaction rate from non-adiabatic transition state theory (NA-TST) is given by \cite{lykhin2016nonadiabatic}
\begin{align}
    k_{\mathrm{NA-TST}} = \frac{ Z_{\mathrm{X}} }{h Z_{\mathrm{R}} } \int_{0}^{\infty} d\varepsilon \mathspace P(\varepsilon) \ \mathrm{exp}(-\beta \varepsilon  ). \label{eq:k_nats}
\end{align}
Here $\beta = 1 / ( k_{\mathrm{B}} T )$, $Z_{\mathrm{X}}$ and $Z_{\mathrm{R}}$ are the rovibrational partition functions at the crossing point and reactant geometry, respectively, and $P(\varepsilon)$ is the probability of transitioning between the two electronic states at energy $\varepsilon$. $P(\varepsilon)$ can be computed within the Wentzel-Kramers-Brillouin (WKB) approximation 
\cite{nikitin1974theory, delos1973reactions}. The result is \cite{liu2014modeling}
\begin{align}
    & k_{\mathrm{NA-TST}} = k_{\mathrm{ISC}} \mathspace \mathrm{exp}(-\beta \Delta G^{X}),  \label{eq:k_na_tst}
\end{align}
where the intersystem crossing rate is
\begin{align}
    & k_{\mathrm{ISC}} = \frac{\pi^{3/2} \alpha}{ 2 h \sqrt{\lambda / (k_{\mathrm{B}} T) } } \left[ 1 + \frac{1}{2} \mathspace \mathrm{exp} \left( \frac{1}{12 \mathspace \alpha^2 \mathspace ( k_{\mathrm{B}} T \lambda )^3 } \right) \right], \label{eq:k_isc}
\end{align}
and
\begin{align}
    & \alpha = \frac{4 H_{\mathrm{SO}}^{3/2} }{\hbar } \left( \frac{\mu}{F_g \vert \Delta \mathbf{F} \vert  } \right)^{1/2} \\
    & \lambda = \frac{\vert \Delta \mathbf{F} \vert}{2 F_g H_{\mathrm{SO}} } \\
    & F_g = \bigg\vert \sum_{j=1}^{N} \sum_{n=1}^{3} \left(\mathbf{F}_{S_0}\right)_{jn} \left(\mathbf{F}_{T_1}\right)_{jn}  \bigg\vert^{1/2} .
    \label{eq:k_isc2} 
\end{align}
Here $H_{\mathrm{SO}}$ is the spin-orbit coupling, $\mu$ is the reduced mass of the reaction coordinate, $F_g$ is the geometric mean of the singlet and triplet forces at the crossing, $N$ is the number of atoms, and $\Delta \mathbf{F}$ is the force difference (Eq. (\ref{eq:df})). In NA-TST, the reaction coordinate is the direction of $\Delta \mathbf{F}$. The reduced mass is then given by
\begin{align}
    \mu = \left(\frac{1}{\vert \Delta \mathbf{F} \vert^2 } \sum_{j=1}^{N} \sum_{n=1}^{3} \Delta \mathbf{F}_{jn}^{2}  m_j^{-1} \right)^{-1},
\end{align}
where $m_j$ is the mass of the $j^{\mathrm{th}}$ atom. Notice that Eq. (\ref{eq:k_isc}) scales quadratically with $H_{\mathrm{SO}}$, as expected from Fermi's golden rule \cite{fermi1995notes}. In re-expressing the equation from Ref. \cite{liu2014modeling} we have used the relation $G_{\mathrm{rv}} = - k_{\mathrm{B}} T \mathspace  \mathrm{log} Z_{\mathrm{rv}}  $ for the rovibrational free energy, and written $G^{\mathrm{X}}= E^\mathrm{X} + G_{\mathrm{rv}}^\mathrm{X}$. We have also corrected two typos (Eq. (7) in Ref. \cite{liu2014modeling} should not contain $k_{\mathrm{B}} T$, and $F_g$ involves a square root rather than a square).

In this work we have approximated $H_{\mathrm{SO}}$ as constant among all azobenzene derivatives. We tested this approximation by computing $H_{\mathrm{SO}}$ at the rotational TS of several derivatives. We included a derivative with chlorine, since, as a relatively heavy third row element, it could increase the coupling. Couplings are available in Q-Chem for TDFFT, but not for SF-TDDFT or CASPT2. They are also not available for CASPT2 in Orca. We therefore used TDDFT with both the B3LYP and BHHLYP functionals, and found that $H_{\mathrm{SO}} \approx 40 \ \mathrm{cm}^{-1}$, with a relative range of 25\% between the smallest and biggest values. The coupling is within a factor of two of the (14, 12) CASPT2 result \cite{cembran2004mechanism}, and the small variation indicates that $H_{\mathrm{SO}} \approx \mathrm{const}.$ is a good approximation. It also suggests that the couplings from CASPT2 would be close to constant. In our calculations we therefore set $H_{\mathrm{SO}} = 20 \ \mathrm{cm}^{-1}$, following the (presumably) more accurate CASPT2 result.

It is informative to estimate the impact of this constant coupling approximation. To do so, we note that the ISC rate scales quadratically with the spin-orbit coupling. Thus a relative change of 25\% leads to a 56\% change in the rate. Converting this to an effective activation entropy, and using $t_{\mathrm{ISC}}$ = 4.2 ps as discussed below, gives a change of 0.26 kcal/mol to $\Delta G^{\mathrm{eff}}$. Thus the maximum error from this approximation is only 0.26 kcal/mol, and is thus negligible compared to other sources of error.

The ISC rate must be multiplied by two for the singlet $\to$ triplet transition, and by three for the triplet $\to$ singlet transition \cite{cembran2004mechanism}. Using Eq. (\ref{eq:k_isc}) together with $H_{\mathrm{SO}} = 20 \ \mathrm{cm}^{-1}$, we found that $t_{\mathrm{ISC}} = 5.7$ ps for $S \to T$ and 4.2 ps for $T \to S$. These are within a factor of two of the results quoted in Ref. \cite{cembran2004mechanism} using Fermi's golden rule, thus validating both their approach and ours. Note that we use the same spin-orbit coupling constant as Ref. \cite{cembran2004mechanism}, and so the differences come solely from the different rate formulas. The benefit of Eq. (\ref{eq:k_isc}) is that it does not contain the Franck-Condon factors used in Fermi's golden rule expressions \cite{cembran2004mechanism, valiev2018first}. Such terms are computationally expensive, and cannot be easily computed for anharmonic modes such as the reaction coordinate. We found that $t_{\mathrm{ISC}}$ was essentially constant among derivatives: the minimum time was 4.8 ps, and the maximum time was 5.8 ps.

\section{Connection between non-adiabatic and Eyring transition state theory}

The intersystem crossing rate in NA-TST can be connected to the experimental $\Delta S^{\dagger}$. Experimentally one starts by assuming an Arrhenius transition rate, 
\begin{align}
k_{\mathrm{A}} = A e^{- E_a/ (k_{\mathrm{B}} T )}. \label{eq:tst_a}
\end{align}
Here $E_a$ is the activation energy and $A$ is the Arrhenius prefactor. The two parameters are obtained experimentally from a plot of $\mathrm{log} \mathspace k$ vs. $1/T$ \cite{rietze2017thermal}. Differentiating $\mathrm{log} \mathspace k$ with respect to $1/T$ in Eqs. (\ref{eq:tst}) and (\ref{eq:tst_a}), equating the two results, and invoking Eq. (\ref{eq:tst}) yields \cite{rietze2017thermal}
\begin{align}
   & \Delta S^{\dagger} = k_{\mathrm{B}} \left( \mathrm{log} \left( A \frac{h}{ k_{\mathrm{B}} T } \right) - 1\right), \label{eq:dS_dagger}
\end{align}
where the slight temperature dependence of $S$ and $H$ has been neglected. If the process is actually mediated by ISC, then we can apply the same logic to Eq. (\ref{eq:k_na_tst}), which yields 
\begin{align}
    & E_a = \Delta H^{\mathrm{X}} + \frac{k_{\mathrm{B}} T}{2} - \frac{3 \eta k_{\mathrm{B}}}{T^2} \frac{\mathrm{e}^{\eta / T^3}}{2 + \mathrm{e}^{\eta / T^3} } \label{eq:Ea_na} \\ 
    & \eta = \frac{1}{12 \alpha^2 k_{\mathrm{B}}^3 \lambda^3 }. 
\end{align}
Inserting Eqs. (\ref{eq:Ea_na}) and (\ref{eq:tst_a}) into (\ref{eq:k_na_tst}) then gives
\begin{align}
    A_{\mathrm{NA-TST}} = k_{\mathrm{ISC}} \mathspace  \mathrm{exp} \left(\frac{\Delta S^{\mathrm{X}}  }{k_{\mathrm{B}}} + \frac{1}{2}  \right) \mathspace \mathrm{exp} \left( \frac{-3 \eta \mathspace e^{\eta / T^3} }{ (2 + e^{\eta /T^3}) T^3 }  \right) \label{eq:A_na}.
\end{align}
Typically $\eta / T^3 \ll 1$, and so the last exponential in Eq. (\ref{eq:A_na}) is of order one. Neglecting this term and inserting the result into Eq. (\ref{eq:dS_dagger}) gives the effective activation entropy,
\begin{align}
    \Delta S^{\mathrm{eff}} = k_{\mathrm{B}} \left(  \mathspace \mathrm{log} \mathspace \frac{h k_{\mathrm{ISC}}}{k_{\mathrm{B}} T}  - \frac{1}{2} \right)+ \Delta S^{\mathrm{X}}. \label{eq:s_app}
\end{align}
This is the entropy that would be inferred from experiment, given a triplet-mediated process with intersystem crossing rate $k_{\mathrm{ISC}}$, and an entropy difference $\Delta S^{\mathrm{X}}$ between the intersection geometry and the reactant geometry. 


\section{Activation free energies by mechanism}
\label{sm_sec:dg_by_mechanism}
\begin{figure*}[t]
    \centering
    \includegraphics[width=\textwidth]{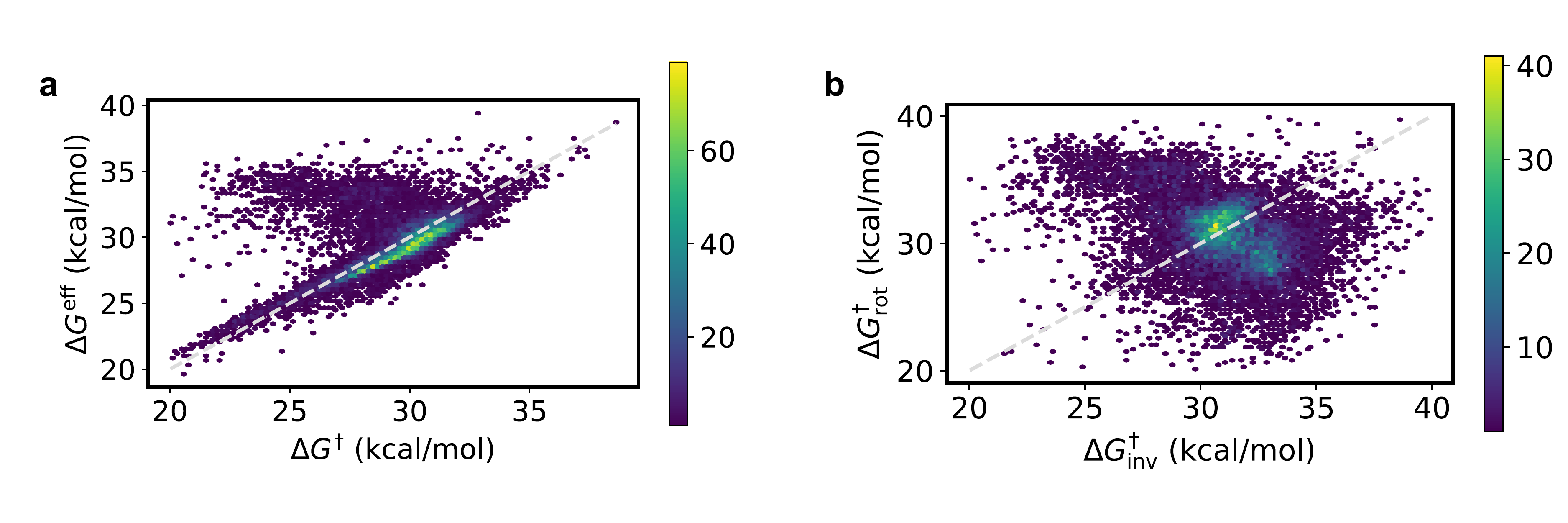}
    \caption{Reaction barriers for different mechanisms. (a) $\Delta G^{\mathrm{eff}}$ from ISC vs. $\Delta G^{\dagger}$ from TST on the $S_0$ surface. (b) $\Delta G^{\dagger}$ from TST for rotation vs. inversion.}
    \label{fig:mech_dg}
\end{figure*}

SI Fig. \ref{fig:mech_dg} compares the activation free energies of the different mechanisms. The plots are for the 19,000 compounds screened in the main text. Panel (a) compares triplet-mediated isomerization with standard $S_0$ isomerization. $\Delta G^{\mathrm{eff}}$ is lower than $\Delta G^{\dagger}$ in 65\% of cases. This indicates that, at the SF-TDDFT level of theory, isomerization proceeds through $T_1$ for roughly 2/3 of the molecules. However, this result should be interpreted with caution. As shown in Table \ref{tab:quantum_chem_methods}, SF-TDDFT overestimates the energy at both the MECP and the TS relative to the highly accurate SF-EOM-CCSD(dT). The overestimate is 5.4 kcal/mol at the MECP and 3.9 kcal/mol at the TS. This means that $\Delta G^{\mathrm{eff}} - \Delta G^{\dagger}$ is overestimated by 1.5 kcal/mol. We therefore expect that, for the 19,000 compounds screened, $\Delta G^{\mathrm{eff}}$ should be lower than $\Delta G^{\dagger}$ in \textit{more} than 65\% of cases. Indeed, if we assume that these overestimates are constant among species, and thus subtract 1.5 kcal/mol from all $\Delta G^{\mathrm{eff}}$, we find that the triplet mechanism is preferred in 88\% of cases. This result should also be interpreted with caution, however, since it neglects errors in $\Delta S^{\mathrm{eff}}$ and $\Delta S^{\dagger}$, and assumes that the energy error is constant among all species.

Panel (b) compares the rotation and inversion mechanisms in TST. $\Delta G^{\dagger}_{\mathrm{rot}}$ is lower than $\Delta G^{\dagger}_{\mathrm{inv}}$ in 55\% of all cases. Interestingly, this means that there is no strong preference for either mechanism at the SF-TDDFT level of theory. This should be contrasted with the species in Fig. \ref{fig:barriers_vs_expt}, for which rotation was preferred in every case.

\section{Thermal isomerization rates}
\begin{figure*}[t]
	\centering
	\includegraphics[width=\textwidth]{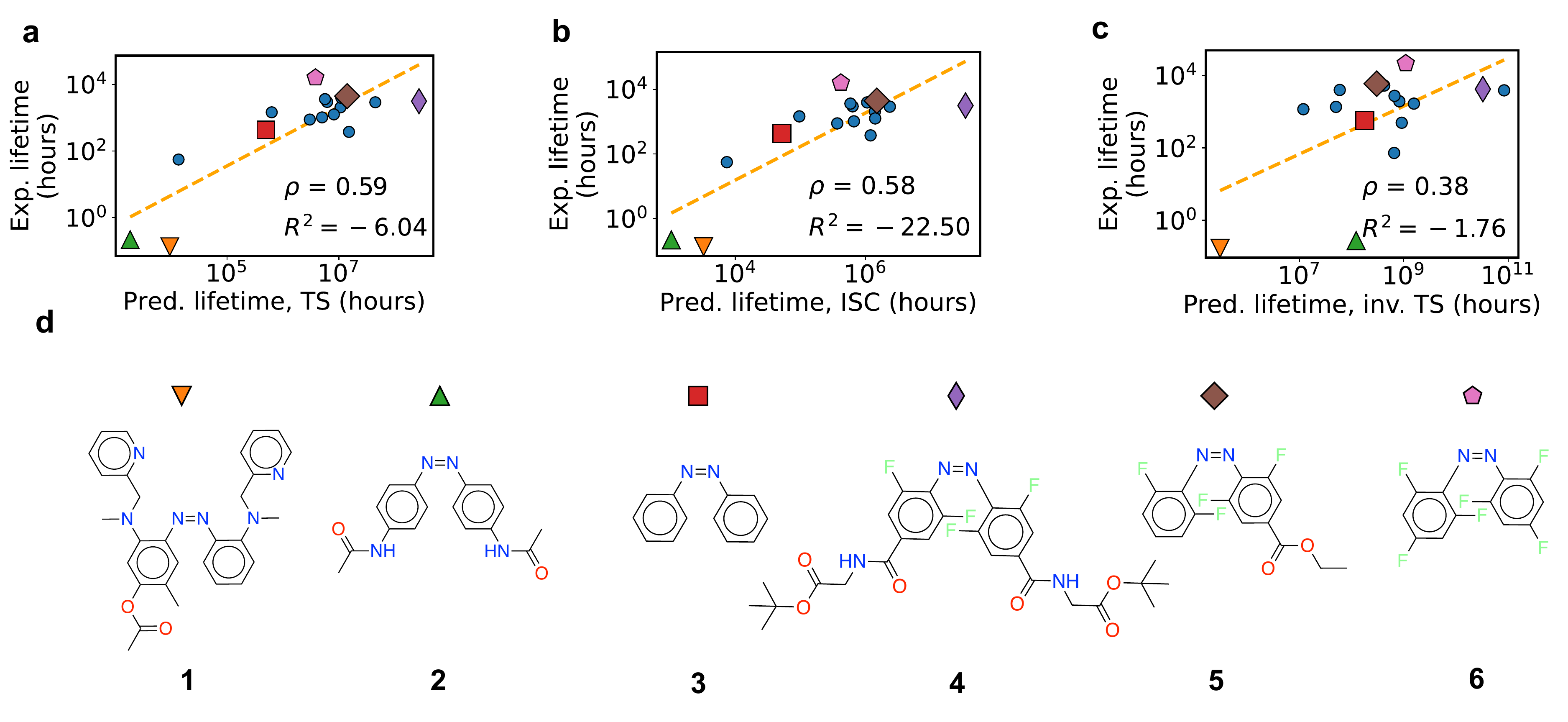}
	\caption{Experimental vs. predicted thermal lifetimes. Dotted orange lines are linear regression results from predicted to experimental $\Delta G$, converted to a lifetime. $\rho$ denotes the Spearman rank correlation. $R^2$ is computed between  the regression results and the experimental data. (a) Prediction accuracy using TS theory. (b) Prediction accuracy using intersystem crossing. (c) As in (a), but with only the inversion mechanism. (d) Selected compounds highlighted in panels (a)-(c).}
	\label{sm_fig:barriers_vs_expt_rates}
\end{figure*}

SI Fig. \ref{sm_fig:barriers_vs_expt_rates} compares predicted and experimental thermal lifetimes, rather than $\Delta G$ values. It contains the same information as Fig. \ref{fig:barriers_vs_expt} in the main text, but with $\Delta G$ converted to a rate using Eq. (\ref{eq:tst}), with $T=298.15$ K. The lifetime is the reciprocal of the rate. Notice that we first converted experimental rates to $\Delta G$, given the temperature of the experiment, and then back to a rate using $T=298.15$ K. This ensures that the lifetimes are all compared at the same temperature. The line of best fit was computed for $\Delta G$, and then converted to a lifetime.

We see that the experimental lifetimes span five orders of magnitude. Species \textbf{1} has the shortest lifetime at 10 minutes, while species \textbf{6} has the longest lifetime at 2.5 years. The systematic overestimation of the energies at the MECP and the TS leads to an overestimate of the lifetimes. The predicted lifetimes are respectively 4, 5, and 7 orders of magnitude too high for TST, ISC, and TST with only inversion. However, since the errors are systematic, the model still has a high Spearman rank correlation with experiment. Converting from $\Delta G$ to lifetime does not affect the ranking, and so the Spearman rank coefficients are the same as in Fig. \ref{fig:barriers_vs_expt}. $R^2$ is negative because of the exponential dependence of lifetime errors on $\Delta G$ errors.

\section{Training}
\subsection{Singlet models}
\label{subsec:training_singlet}

Three different models were pre-trained on 680,736 gas-phase SF-TDDFT/6-31G* calculations from Ref. \cite{axelrod2022excited}. 95\% of the data was used for training, 4\% was used for validation, and 1\% for testing. Each model was initialized with a different random seed and trained on different train/validation/test splits. Each model was then fine-tuned with SF-TDDFT/6-31G* using a C-PCM model of water \cite{truong1995new, barone1998quantum, cossi2003energies}. The final models were trained on 42,938 geometries, with 500 used for validation and 500 for testing. Different splits were used to fine-tune each of the final models. 

The singlet ground-state corresponds to an excitation from a triplet reference state in SF-TDDFT. We identified two singlets as the two states with the lowest $\langle S^2 \rangle $ of the three-lowest energy excitations \cite{axelrod2022excited}. The lower energy singlet was then taken as the ground singlet. Geometries with $\langle S^2 \rangle > 1$ in the ground singlet state were discarded. The spin contamination in the training set was fairly low, with $\mathrm{mean} \langle S^2 \rangle  = 0.16$. 

Approximately 6\% of species in the training set had non-zero charge ($+1$, $+2$, or $+4$). The model does not use charge in the input; however, all charge states in this work can be inferred from the number of bonds for each atom. For example, if nitrogen has a single bond to four different atoms, then it must have a partial charge of $+1$ to have a full octet. Since bonds can be inferred from atom type and distances \cite{grimme2010consistent}, the model can learn the effect of different charge states from positions and atomic numbers alone.

Training was performed over energies and forces/force couplings in units of kcal/mol and kcal/mol/\AA, respectively. Per-species reference energies were subtracted from each energy. These were obtained by summing atomic reference energies, computed using multi-variable linear regression from (atom type, count) to relaxed geometry energy computed with SF-TDDFT in vacuum. The set of vacuum reference energies was used for both vacuum and solvent training.

Configurations with 10-$\sigma$ energy outliers were removed prior to training. Those with forces $\geq$450 kcal/mol/\AA \ from the mean or energies $\geq$900 kcal/mol from the mean were also removed. Geometries were not removed on the basis of $X$-$\sigma$ force deviations, since many geometries were close to TSs and hence had near-zero forces. 

Models were trained with the Adam algorithm, using a batch size of 60 for the gas-phase models and 20 for the fine-tuned models. We used an MAE loss for the forces and energies:
\begin{align}
    \mathcal{L} \ = \ & \frac{1}{M} \mathspace \sum_{i=1}^{M} \rho_E \mathspace \vert E_i - \hat{E}_i   \vert  
    \ + 
    \ \frac{1 }{3 \sum_{i=1}^{M} N_i } \mathspace \sum_{i=1}^{M} \sum_{j=1}^{N_i} \sum_{n=1}^{3} \rho_F \mathspace \vert F_{j, n} - \hat{F}_{j, n}   \vert.
\end{align}
Here $\rho_E = 0.2$ is the energy weight, $\rho_F = 1.0$ is the force weight, $M$ is the number of geometries in a batch, $N_i$ is the number of atoms in geometry $i$, $\hat{X}$ is a predicted quantity and $X$ is its true value. We used an MAE loss instead of a mean-squared-error loss, since this led to significantly better model performance.

The learning rate was initialized to $10^{-4}$ and reduced by a factor of two if the validation loss had not improved in $X$ epochs. $X$ was set to 10 for the gas-phase models and 50 for the fine-tuned models. Training was stopped when the learning rate fell below $10^{-5}$. The final model was selected as the one with the lowest validation loss. Training was performed on a single 32 GB Nvidia Volta V100 GPU, and took approximately 2 days for each of the models.

\subsection{Triplet models}
Triplet models were trained on the $S_0$/$T_1$ gap. They were pre-trained using energies from Ref. \cite{axelrod2022excited}. Gradients were not used since they had not been computed.  For fine-tuning we computed gradients of both the singlet and triplet state for all geometries, and thus used both the gap and its gradient for training. The same loss tradeoff was used as for the singlet model. The learning rate was set $5 \times 10^{-5}$ for pre-training and $10^{-4}$ for fine-tuning; the former was chosen because of instabilities in pre-training. To identify the triplet state, we first identified the two singlet states from the three lowest-energy excitations using the method above. The triplet state was then the remaining one. Data points with $\langle S^2 \rangle < 2.0$ or $\langle S^2 \rangle > 2.4$ were discarded.



SF-TDDFT uses an $M_{\mathrm{S}} = 1$ triplet reference state. The excited $M_{\mathrm{S}} = 0$ triplet and the reference $M_{\mathrm{S}} = 1$ triplet should have the same energy. We have found this to be true for spin-flip coupled-cluster methods, but not for SF-TDDFT. This non-zero energy difference was also observed in Ref. \cite{huix2010assessment}. The authors explained that the difference is due to orbital relaxation, since the orbitals are relaxed for the reference state but not for the TDDFT states.

It is not immediately clear which triplet energy should be used for ISC. We tested results using the reference triplet, the excited triplet, and an average of the two. We compared $E^{\mathrm{X}}$ to that of spin-flip coupled cluster for azobenzene (SI Sec. \ref{sec:benchmark}), and found that the best agreement was for the triplet average. The excited triplet overestimated $E^{\mathrm{X}}$ by 5 kcal/mol, and the reference triplet underestimated it by nearly the same amount. However, we found that the correlation with experiment was slightly better when we used the SF-TDDFT triplet. Further, this is more consistent with the principle of treating singlet and triplet states on ``equal footing''. We therefore trained our models on the excited-state triplet energies. The same network architecture and training parameters were used as for the singlet models.

\subsection{Singlet excitation model}
The singlet excitation model was trained directly on the $S_0$/$S_1$ gap. It was first pre-trained using energies and gradients from Ref. \cite{axelrod2022excited}. It was then fine-tuned using energies only, since we did not compute $S_1$ gradients on the new geometries.

\section{Filtering the screening results}
\label{sm_sec:filtering}
We only kept the species that satisfied the following conditions:
\begin{enumerate}
	\item \textbf{Endpoints are done}. Both \textit{cis} and \textit{trans} must have conformer ensembles and a minimum-energy geometry with no imaginary frequencies.
	\item \textbf{Endpoints maintain \textit{cis}/\textit{trans} isomerism}. The optimized \textit{cis} and \textit{trans} isomers must actually be \textit{cis} and \textit{trans}, respectively. This is checked with RDKit.
	\item \textbf{TS conformer generation is finished for each mechanism}. A TS conformer ensemble must be generated for each of the four mechanisms.
	\item \textbf{At least one TS is converged for each mechanism}.  To be converged, the TS must have only one imaginary frequency, and its magnitude must exceed 200 $\mathrm{cm}^{-1}$.
	\item \textbf{Graph is unchanged}. All of the TS and endpoint geometries must have the same molecular graph as the input SMILES. This is checked using D3 coordination numbers. Atom pairs with coordination numbers $\geq 0.95$ are assigned a bond. The resulting bond list is checked against that of the original SMILES string.
	\item \textbf{Vibrational frequencies are reliable}. A frequency calculation is deemed reliable if $\vert \Delta G^{\dagger}  - \Delta E^{\dagger} \vert \leq 10$ kcal/mol.
	\item \textbf{For the two singlet-triplet MECPs, one is closer to \textit{cis} and the other to \textit{trans}}. This is checked using the root-mean-square displacement of the CNNC atoms after alignment.  \label{item:proper_isc}
	\item \textbf{Each rotational TS connects to \textit{cis} and \textit{trans} through the IRC}. This is true if condition \ref{item:proper_isc} is satisfied.
\end{enumerate}
25,000 species were screened in total, and 18,877 remained after applying these filters.

\section{Experimental data}
\label{sec:si_experiment}
Most papers reported either the thermal half-life $\tau_{1/2}$, or the thermal lifetime $\tau$. These are related through $\tau = (e/2) \mathspace \tau_{1/2} $. In some cases the authors did not provide $\tau$ or $\tau_{1/2}$, but did plot the absorption spectrum at different time intervals. In these cases we inferred the thermal lifetime from the absorption plots.

In all cases we computed the activation free energy from Eyring TST using $\tau$:
\begin{align}
	\Delta G^{\dagger} = k_{\mathrm{B}} T \mathspace \mathrm{log} \frac{ k_{\mathrm{B}} T \tau  }{ h }.
\end{align}
Note that different works used different temperatures, and so $T$ depends on the source.

\section{Note on mechanisms}
\subsection{Concerted inversion}
\label{subsec:concerted}
 In our own calculations we have found that concerted inversion has two imaginary frequencies, and therefore is not a true TS. This is because it is a combination of two different inversions, each of which has one imaginary frequency. It is therefore higher in energy than a single inversion, not a true TS, and can be excluded from possible thermal mechanisms.

\subsection{Inversion-assisted rotation}
\label{subsec:inversion}
A relaxed scan from $\alpha \approx 120^{\circ}$ to $\alpha \approx 180^{\circ}$ will either end in a pure inversion TS or an inversion-assisted rotation TS, depending on the substituents. One can also be converted into the other through a conformational search with fixed CNNC atoms. Therefore, we grouped inversion and inversion-assisted rotation together.

\section{Quantum chemistry benchmark}
\label{sec:benchmark}
\begin{table*}[t]
\small
\centering
\begin{tabular}{c|c|c|c|c | c | c }
     \hline
     Method & Basis & Restricted? & \begin{tabular}{@{}c@{}}Treatment of \\ Static Correlation \end{tabular} & \begin{tabular}{@{}c@{}}Treatment of \\ Dynamic Correlation \end{tabular} & $\Delta E^{\dagger}$ & $\Delta E^{\mathrm{X}}$   \\
     \hline
     DFT & cc-pVDZ &  Yes & Poor & Strong & \ 35.6 \ & n/a \\
     DLPNO-UCCSD(T) & F12-cc-pV(D, T)Z extrap. &  No & Poor & Very strong & 42.1 & n/a \\
     SF-TDDFT & 6-31G* &  No & Moderate & Strong & 31.5 & 27.8 \\
     SF-TDDFT (D3BJ) & 6-31G* &  No & Moderate & Strong & 32.6 & 28.9   \\
     CASPT2 (14, 12)  & cc-pVDZ &  Yes & Strong & Moderate & 25.4 & 18.7 \\
     SF-EOM-CCSD(dT) & 6-31G* & Yes & Moderate  & Very strong  & 28.7 & \ 23.5 \ \\
    \hline
\end{tabular}
\caption{Summary of the methods and results from our benchmark. $\Delta E^{\dagger}$ is the energy difference between the TS and the reactant, and $\Delta E^{\mathrm{X}}$ is the energy difference between the singlet-triplet crossing and the reactant. They are shown schematically in Fig. \ref{fig:isc} in the main text. DFT uses the B3LYP functional with D3BJ dispersion, while SF-TDDFT uses the BHHLYP functional, both with and without D3BJ dispersion. Energy differences are given in kcal/mol.}
\label{tab:quantum_chem_methods}
\end{table*}

Here we compare the torsional energy profiles of unsubstituted azobenzene predicted by different electronic structure methods. We analyze results from DFT, DLPNO-UCCSD(T) (domain based local pair-natural orbitals \cite{neese2009efficient, neese2009accurate, neese2009efficient2, liakos2011weak, hansen2011efficient, riplinger2013efficient, riplinger2013natural, riplinger2016sparse, datta2016analytic, saitow2017new} for unrestricted coupled cluster with single, double, and perturbative triple excitations), SF-TDDFT, CASPT2 (complete active space with second-order perturbation theory), and SF-EOM-CCSD(dT) (spin-flip equation-of-motion coupled-cluster with single, double, and perturbative triple excitations \cite{manohar2008noniterative}). The methods are summarized in Table \ref{tab:quantum_chem_methods}. 
For DFT we use the B3LYP functional \cite{becke1993density} with D3 \cite{grimme2010consistent} Becke-Johnson (BJ) dispersion \cite{grimme2011effect}. For SF-TDDFT we use the BHHLYP functional both with and without D3BJ dispersion. 

For B3LYP-D3BJ we use the double-zeta correlation-consistent (cc) basis set cc-pVDZ \cite{dunning1989gaussian}. This model chemistry predicts activation free energies of azoarene isomerization that are in good agreement with experiment \cite{adrion2021benchmarking}. For DLPNO-UCCSD(T) we extrapolate to the complete basis-set limit using explicit correlation (F12) \cite{pavovsevic2017sparsemaps} with double- and triple-zeta basis sets (cc-pV(D, T)Z). Such high-accuracy basis sets can be used because of the computational efficiency of DLPNO. For BHHLYP SF-TDDFT we use the 6-31G* basis \cite{hehre1986p}, since this is a medium-cost, medium-accuracy basis set that can be used to generate large amounts of data for ML. Similar quality basis-sets are used for CASPT2 and SF-EOM-CCSD(dT) due to computational constraints. 
B3LYP-D3BJ, CASPT2, and DLPNO-CCSD(T) calculations are performed with Orca 5.2 \cite{neese2020orca}. For DFT and CCSD(T) calculations in Orca we use the resolution of identity \cite{whitten1973coulombic,baerends1973self, dunlap1979some,van1988ab,kendall1997impact, eichkorn1995auxiliary, eichkorn1997auxiliary} and chain-of-spheres \cite{neese2009efficientcosx} approximations (\texttt{RIJCOSX}). The \texttt{TightPNO} setting is used for DLPNO calculations. All other calculations are performed with Q-Chem 5.3.  

\begin{figure*}[t]
    \centering
    \includegraphics[width=\textwidth]{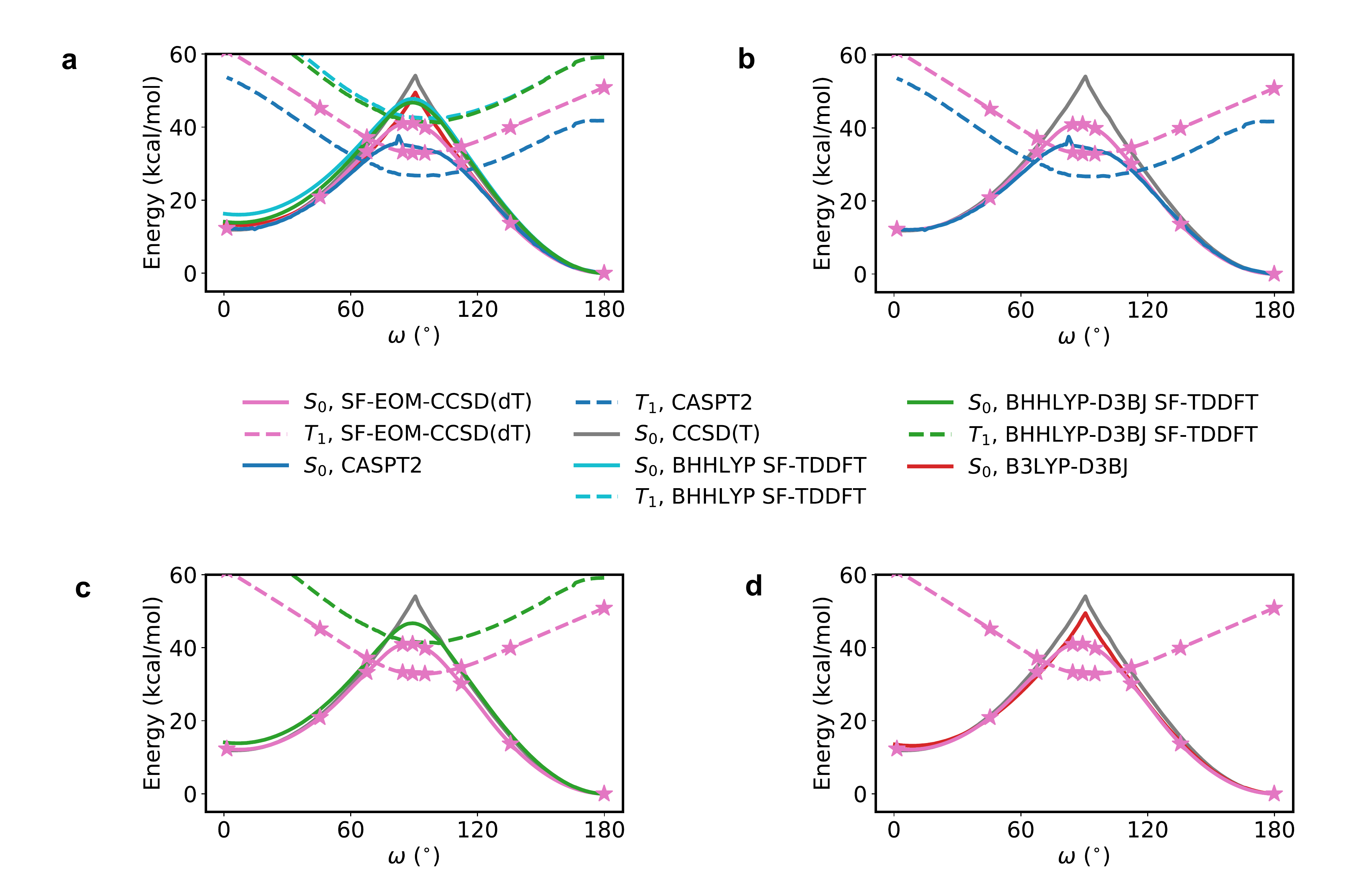}
    \caption{Torsional energy profile of azobenzene computed with different methods. Singlet and triplet energies are shown with full and dashed lines, respectively. Singlet and triplet energies computed with the same method are shown with the same color. (a) Results from all methods. For ease of visualization, different subsets of methods are compared in panels (b)-(d). (b) The multi-reference methods SF-EOM-CCSD(dT) and CASPT2 are compared with the single-reference method CCSD(T). (c) The multi-reference methods SF-EOM-CCSD(dT) and SF-TDDFT are compared with the single-reference method CCSD(T). (d) The multi-reference method SF-EOM-CCSD(dT) is compared with the single-reference methods B3LYP and CCSD(T).  }
    \label{fig:benchmark}
\end{figure*}
We performed a relaxed scan from $\omega=0$ to $\omega=180^{\circ}$ in $1^{\circ}$ steps using our trained NN potential. We constrained the dihedral angle at each step using a force constant of 1 Ha.  We then performed single-point gas-phase calculations with the methods given above. Calculations were performed on all geometries, except for SF-EOM-CCSD(dT), which was performed on only nine geometries due to high computational cost. Those points are shown as stars, and are connected by a smooth line using a quadratic interpolation. Triplet calculations with DLPNO-CCSD(T) failed to converge and are hence not reported.

Results are shown in Fig. \ref{fig:benchmark}. The plots contain several salient features. First, the single-reference methods DFT and UCCSD(T) predict a cusp near $\omega = 90^{\circ}$, while the multi-reference methods produce a smooth maximum of lower energy. This can be seen in Fig. \ref{fig:benchmark}(d), where DFT (red) closely tracks UCCSD(T) (gray) until $\omega=90^{\circ}$. Both methods have a cusp at $\omega=90^{\circ}$, but the energy from DFT is lower. Similar results have been found for ethylene torsion \cite{shao2003spin}. 

As shown in Table \ref{tab:quantum_chem_methods}, DFT gives an activation energy of 35.6 kcal/mol, while UCCSD(T) gives 42.1 kcal/mol. The cusp in the DFT energy explains why previous work could not optimize the rotational TS with DFT \cite{rietze2017thermal}. Optimization methods such as EVF \cite{baker1986algorithm} use a quadratic expansion of the PES to locate critical points. This expansion fails in the vicinity of a cusp, since the Hessian is undefined at this point. 

Previous calculations with DFT found that inversion is the preferred mechanism in gas phase \cite{dokic2009quantum}. However, given that the DFT activation energy is 10 kcal/mol higher than CASPT2, such conclusions are not to be trusted. Indeed, the inversion activation energy was found to be 31 kcal/mol with CASPT2 (10, 8) \cite{casellas2016excited}, which is 6 kcal/mol higher than the CASPT2 rotation barrier here. Thus according to CASPT2, isomerization would proceed by rotation even if it were not mediated by ISC.

It is also interesting that the maximum value of the CCSD(T) $T_1$-diagnostic \cite{lee1989diagnostic} is 0.013, which occurs at $\omega=90^{\circ}$. The minimum value is 0.01. A value above 0.02 indicates that the problem has multi-reference character, and that CCSD(T) should not be trusted. The diagnostic stays well below 0.02 despite the clear multi-reference nature of the problem. We also note that the unrestricted CCSD(T) results collapsed to the restricted results, with the square spin $\langle S^2 \rangle = 0$ for all geometries.

A second important result is that all multi-reference methods predict smooth TS maxima with singlet-triplet crossings on either side. Examples are CASPT2 and SF-EOM-CCSD(dT) in Fig. \ref{fig:benchmark}(b), and  SF-EOM-CCSD(dT) and SF-TDDFT in Fig. \ref{fig:benchmark}(c). However, the $\Delta E^{\dagger}$ and $\Delta E^{\mathrm{X}}$  predicted by the different methods are in quantitative disagreement. The activation energies predicted by CASPT2, SF-EOM-CCSD(dT), and SF-TDDFT (D3BJ) are 25.4, 28.7, and 32.6 kcal/mol, respectively. The singlet-triplet crossing energies are 18.7, 23.5, and 28.9 kcal/mol respectively. Given the disagreement between CASPT2 and SF-EOM, it is not immediately clear which method should be taken as the ``ground truth''. 

On balance, however, it seems that SF-EOM-CCSD(dT) is likely more accurate. One reason is that its predictions are in quantitative agreement with MR-CISD+Q (multi-reference configuration interaction with singles and doubles, plus a Davidson correction) for a standard rotational conical intersection benchmark \cite{gozem2013conical}. Its treatment of static correlation should therefore be sufficient for azobenzene, while its treatment of dynamic correlation is better than that of CASPT2. Moreover, its predicted activation enthalpy is in better agreement with experiment. The experimental activation enthalpy is 21.1 kcal/mol \cite{asano1981temperature}, which corresponds to $\Delta H^{\mathrm{X}}$ in the ISC picture. Our NN calculations indicate that $\Delta H^{\mathrm{X}} \approx \Delta E^{\mathrm{X}} - 2.5$ kcal/mol for azobenzene. Therefore, $\Delta H^{\mathrm{X}} \approx 21.0$ kcal/mol with SF-EOM-CCSD(dT), while $\Delta H^{\mathrm{X}} \approx 16.2$ kcal/mol with CASPT2. The former agrees better with experiment than the latter.

A third important result is that SF-TDDFT's predictions are in qualitative agreement with other multi-reference methods, but not quantitative agreement. As in the other methods, SF-TDDFT predicts a smooth maximium with singlet-triplet crossings on either side of the TS. However, its predicted barrier is 3.9 kcal/mol higher than that of SF-EOM, and 7.2 kcal/mol higher than that of CASPT2. Its singlet-triplet crossing energy is 5.4 kcal/mol higher than SF-EOM, and 10.2 kcal/mol higher than CASPT2. These errors make it difficult to compare the ISC reaction rate with the Eyring reaction rate using SF-TDDFT. Hence in the main text we simply assume that all azobenzene derivatives isomerize through ISC.

Lastly, the plots show the importance of adding dispersion corrections to DFT. Figure \ref{fig:benchmark}(a) includes results of SF-TDDFT with and without dispersion. These are shown in green and cyan, respectively. When dispersion is included, all single- and multi-reference methods are in good agreement away from $\omega = 90^{\circ}$. When dispersion is not included, SF-TDDFT overestimates the \textit{cis} energy, because the attractive force between the two benzene rings is underestimated. Similar conclusions have been made in previous work \cite{rietze2017thermal, adrion2021benchmarking}. There is still some residual error in the \textit{cis} energy, even with dispersion, which does not occur with B3LYP-D3BJ. This is likely because the D3 parameters for BHHLYP were chosen for ground-state DFT, not SF-TDDFT.




\bibliography{main}

\end{document}

%% file: preamble.tex
\usepackage{amsthm}
\usepackage{mathtools}
\usepackage{physics}
\usepackage{xcolor}
\usepackage{graphicx}
\usepackage{graphics}
\usepackage[fontsize=10]{fontsize}
\usepackage{amsmath}
\usepackage{hyperref}

\newcommand{\mathspace}{\hspace*{0.07cm}}

 \SectionNumbersOn